\documentclass[prb,groupedaddress,showpacs,twocolumn]{revtex4}
\usepackage{graphicx}
\usepackage{amsmath}
\usepackage{amssymb}

\newcommand{\fref}[1]{Fig.~\ref{#1}}
\newcommand{\sref}[1]{Sec.~\ref{#1}}
\newcommand{\aref}[1]{Appendix~\ref{#1}}

\newcommand{\normwidth}{0.7\columnwidth}
\newcommand{\smallwidth}{0.5\columnwidth}

\newcommand{\csr}{\Omega_L^{\rm sr}}

\begin{document}
\title{Zero Field Wigner crystal}
\author{R. Chitra}
\email{chitra@lptl.jussieu.fr} \affiliation{Laboratoire de Physique Theorique des Liquides, UMR 7600, Universite de
Pierre et Marie Curie, Jussieu, Paris-75005, France }
\author{T. Giamarchi}
\email{Thierry.Giamarchi@physics.unige.ch} \affiliation{University
of Geneva, 24 Quai Ernest Ansermet, 1211 Geneva, Switzerland}

\date{\today}

\begin{abstract}
A candidate for the insulating phase of the 2D electron gas, seen
in high mobility 2D MOSFETS and heterojunctions, is a Wigner
crystal pinned by the incipient disorder. With this in view, we
study the effect of collective pinning on the physical properties
of the crystal formed in zero external magnetic field. We use an
elastic theory to describe to long wavelength modes of the
crystal. The disorder is treated using the standard Gaussian
variational method. We calculate various physical properties of
the system with particular emphasis on their density dependence.
We revisit the problem of compressibility in this system  and
present results for the compressibility  obtained  via  effective capacitance measurements  in
experiments using bilayers.  We present results for
the dynamical conductivity, surface acoustic wave anomalies and
the power radiated by the crystal through phonon emission at
finite temperatures.
\end{abstract}


\maketitle

\section{Introduction}

The question of the combined effects of disorder and interactions
in correlated electron systems is one of the more important issues
in solid state physics. The interest in this longstanding problem
has been  revived by the spectacular experimental results on the
two dimensional electron gas\cite{abrahams_review_mit_2d}. From a
theoretical standpoint, both disorder and interactions are
challenges in  themselves. By starting from the noninteracting
limit, considerable progress was achieved in the understanding of
the effects engendered by disorder in the system. The rationale
behind such a limit stems from the expectation that not too strong
interactions will lead to a Fermi liquid behavior, at least in
high dimensions, and therefore, excitations of the system will
behave as free fermions. For non interacting electrons in one and
two dimensions, disorder is known to lead to  the phenomenon of
Anderson localization  where, all states are localized
\cite{anderson_localisation, abrahams_loc,lee_mit_long}. Going
from this non interacting limit to the interacting system has
proven quite challenging. In three or higher dimensions, where Fermi
liquid theory is expected to hold in the absence of disorder, a
renormalization group study
\cite{finkelstein_localization_interactions,belitz_localization_review}
indicated that disorder strongly enhanced the interactions hence
pushing the system away from the noninteracting point. In low
dimensions, the situation is even more complicated because in this
case, even in the absence of disorder, the interactions themselves
have singular effects on the system, and can  lead to non fermi
liquid phases. For example, in  one dimension, where the
interacting system is known to be a Luttinger
liquid\cite{gogolin_1dbook,giamarchi_book_1d}, the combined effect
of disorder and interactions have been shown to lead to behaviors
quite different from that anticipated for a noninteracting
system\cite{giamarchi_loc}.

Two dimensions is thus from this point
of view doubly marginal: first, the effect of interactions even in
the pure system is far from being completely settled and secondly,
the disorder, although it leads to localization for the
noninteracting system does it only marginally. It is therefore,
natural to expect that interactions can alter this picture of
marginally localized states. Consequently, the nature of the phase
diagram of the two dimensional electron gas as a function of the
disorder strength and interactions has been a subject of intense
debate.  There is however, one limit where the combined effect of
interaction and disorder is amenable to an analytical study. This
is the limit of strong interactions, where,  the electrons in the
pure system localize around the sites of a triangular lattice to
form a Wigner crystal\cite{wigner_crystal,ceperley_qmc_wigner}. In
this crystal, and far enough from  melting, the particles become
discernible by their position. Thus quantum effects are simple and
manifest themselves in the quantification of the vibrations of the
electronic crystal. When a disorder potential is added, its effect
is to pin the elastic crystal\cite{fukuyama_cdw_magnetic}. The
pinning of  elastic structures, both classical \cite{blatter_vortex_review,nattermann_vortex_review,giamarchi_vortex_review}
and quantum\cite{giamarchi_quantum_pinning,giamarchi_varenna_wigner_review},  by disorder has been the subject
of intense studies and various methods have been developed to tackle these problems.
These tools have been used to study the magnetic field induced
Wigner crystal formation in the 2D electron
gas\cite{normand_millis_wigner,chitra_wigner_hall,yi_pinning_wigner,chitra_wigner_long,fogler_pinning_wigner}.
The optical conductivity and Hall
coefficients\cite{chitra_wigner_hall,chitra_wigner_long} were
computed and found to be in excellent qualitative agreement with
the experimental
observations\cite{li_conductivity_wigner_magneticfield,li_conductivity_wigner_density}.
Combined with the theoretical analysis, the optical conductivity
provided clear evidence that for certain filling fractions, the
phase realized in this system was indeed a Wigner crystal, weakly
pinned on impurities.

The  formalism used to study the strong field crystal can be
easily modified to obtain the properties of the $B=0$ crystal.
This is of importance to experiments on ultra clean Ga-As
heterojunctions and $Si$ MOSFETS which exhibit the zero field
metal-insulator transition. A better way
to parametrize the system is through the dimensionless parameter
$r_s = a/a_B$, where the lattice spacing $a$ is related to the
electron density $n=1/\pi a^2$ and $a_B$ is the Bohr radius of the
electron in the 2D electron gas. One of the important questions is
whether the insulating phase seen in these systems at low
densities   or large $r_s$ is indeed a disordered Wigner crystal.
 In the present paper we
focus on analyzing the physical properties of  a putative  Wigner
crystal phase. Our approach complements  the standard approach
starting from the noninteracting limit. We investigate the effects
of the disorder on such a system and compute various observable
quantities.

The plan of the paper is as follows. In \sref{sec:physical} we
first use a variational wave function to estimate the effective
particle size i.e., size of the localized wave packet at a site.
This parameter is expected to play an important role as will be
detailed later in the paper. We then present  the model and the
variational solution. The remainder of the paper is devoted to the
computation of various observable quantities: \sref{sec:comp}
deals with the compressibility, \sref{sec:transport} with the
transport properties, in particular, the conductivity,  surface acoustic wave
absorption and  the power radiated by the
pinned crystal, followed by a concluding section \sref{sec:data}.
Technical details have been relegated ton the Appendices.

\section{Model and method}\label{sec:physical}

We will follow the same procedure outlined in
Ref~\onlinecite{chitra_wigner_hall,chitra_wigner_long}. In the
crystal state, the wavefunction of the particles is localized
around positions ${\bf R}_i$. The wave function has a
characteristic width $\xi$ centered around this site.  Since the
localized wavefunction renders  the particles  discernible by
their position,  the crystal  can be considered as a collection of
discernible particles of size $\xi$ that are labelled by their
position ${\bf R}_i$. Then only the (quantized) vibrational modes
corresponding to the displacements of these particles needs to be
retained in any low energy description of the system. Of course,
this greatly simplifies the analysis and is the key ingredient
which permits a resolution of   this problem. As pointed out
previously\cite{chitra_wigner_hall,chitra_wigner_long}, there are
three important lengthscales to describe the crystal: the lattice
spacing $a$, the size $\xi$ of the particles, and the correlation
length of the disorder $r_f$ which we will come back to  later.
For the case of a strong magnetic field the size of the particle
is essentially the cyclotron
radius\cite{chitra_wigner_hall,chitra_wigner_long}. However, for
the $B=0$ Wigner crystal, determining the size of the particle
$\xi$ is a question in itself since it  is determined by the
competition between kinetic energy and interactions.

\subsection{Wavefunction in the crystal state} \label{sec:varxi}

In this section, we use a variational wave function to determine
this effective width $\xi$ as a function of the electron density.
We choose the ground state to be a Slater determinant of gaussian
wave packets, whose width $\xi$ is chosen to be the variational
parameter. The single particle gaussian wave packets centered
around a site $i$ with coordinate ${\bf R}_i$, take the form
$\psi_i({\bf r})= \sqrt{\frac{2}{\xi}}\exp-\frac{{({\bf r} - {\bf
R}_i)}^2}{\xi^2}$. In the absence of a magnetic field, the total
variational energy per site is given by
\begin{equation}
 E(\xi) = \frac{\hbar^2}{m \xi^2} + \frac{e^2}{2\epsilon\xi}\sum_i V({\bf R}_i)
\end{equation}
where the first term  is the kinetic energy of the electrons, and
the second term  is the potential energy arising from coulomb
repulsion and
\begin{equation}
 V({\bf R}_i)= \exp-{\frac{{\bf R}_i^2}{4 \xi^2}}
 I_0(\frac{{\bf R}_i^2}{4 \xi^2}) -\exp-{\frac{{\bf R}_i^2}{8 \xi^2}}
\end{equation}
$I_0$ is a modified Bessel function of the first kind and  ${\bf
R}_i$ denotes the lattice sites. $m$, $e$ and  $\epsilon$ denote
the mass, charge and dielectric constant respectively. Minimizing
the energy $E$ with respect to $\xi$ results in a self-consistent
equation for $\xi$. Excepting the limit of low densities, where
one can obtain an analytic expression for $\xi$, this variational
equation is rather complicated to solve analytically for arbitrary
densities, thereby requiring the use of numerical techniques. In this
paper,  we discuss
only the low density limit or equivalently the deep crystalline phase. 
In this limit, as the
density is lowered, the effective distance between neighboring
sites increases as $a = \sqrt{1/\pi n}$ and  the effective width
of the wave packet increases in manner such that $a/\xi \gg 1$.
Therefore, it is sufficient to retain only the term ${\bf R}_i^2 =
a^2$ in the expression for $V$. Expanding the exponential in
$\xi/a$, we obtain following expression for the energy for low
densities:
\begin{equation} \label{en-lowd}
 E= \frac{\hbar^2}{m\xi^2} + C \frac{e^2\xi^2}{\epsilon a^3} + {\rm \ldots}
\end{equation}
The constant $C$
depends on the coordination number of the lattice and the number
of terms retained in the sum.  Retaining only nearest neighbor
sites on a triangular lattice, we obtain $C\simeq 1.88$.
Minimization of $E$ with respect to $\xi$, yields the result,
\begin{equation}\label{size}
 \xi = \left(\frac{a_{B}}{C}\right)^{\frac{1}{4}}a^{\frac{3}{4}} \equiv n^{{-3/8}}
\end{equation}
$a_B= \frac{\epsilon \hbar^2}{m e^2}$ is the Bohr radius of the
electron. Note that despite the fact that $\xi$  \emph{increases}
with decreasing density, the ratio $\xi/a \sim n^{\frac18} $
\emph{decreases} with decreasing density in accord with the
assumptions made above, thereby justifying the expansion
(\ref{en-lowd}). We emphasize that this result is valid only deep
in the crystalline phase. Close to melting, exchange terms become
important and we need to solve the equation with the full $V$ to
obtain a reasonable variation of the effective particle size with
density. Moreover, when exchange terms  become non-negligible,
the very hypothesis of having discernable particles collapses, and an
effective theory  different from the one presented in the ensuing section
is required.

In the presence of a magnetic field, the same calculation can be
done to obtain $\xi$ as a function of the density and the field.
For the purely gaussian wave packet, the magnetic field
contributes a term $m\omega_c^2 \xi^2/2$ to the energy per site.
In the limit of ultra strong magnetic fields dominating  the
coulomb repulsion, we recover the result that $\xi =
\sqrt{\frac{\hbar}{eB}}$  which is just the cyclotron length. For
arbitrary fields,
\begin{equation}
 \xi=  \left[\frac{\hbar^2}{ m^2 \omega_c^2 + C m e^2 / \epsilon a^3}\right]^{\frac14}
\end{equation}
However, the very assumption of a gaussian wave packet implies
that the electrons are confined to the lowest Landau level and
hence the above result holds, strictly speaking, only for strong
magnetic fields, and should be viewed as a convenient
interpolation formula.

\subsection{Elastic Hamiltonian } \label{sec:model}

Now that we have the important parameters characterizing the
crystal, we  present  the elastic Hamiltonian describing the
crystal phase. We use the same recipe as outlined in
Ref.~\onlinecite{chitra_wigner_hall,chitra_wigner_long}, and
recall only the main steps here. In the crystalline phase, the
electrons occupy the sites of a triangular lattice with a lattice
constant $a$ which is related to the density of electrons by $n
\sim (\pi a^2)^{{-1}}$. A particle at a site $i$ is displaced from
its mean equilibrium position denoted by ${\bf R}_i$, by ${\bf
u}({\bf R}_i,t)$. In the continuum limit, the vibration modes of
the crystal lead to the following elastic action in Fourier space
\begin{widetext}
\begin{multline}\label{eq:ham}
 S = \frac1{2\beta} \sum_{\omega_n}\int d^2q  \left[
 u_{T}({\bf q},\omega_n)(\rho_m\omega_n^2 +\Omega_T({\bf q}))u_{T}(-{\bf
 q},-\omega_n)
 + u_{L}({\bf q},\omega_n)(\rho_m\omega_n^2 +\Omega_L({\bf q}))u_{L}(-{\bf q},-\omega_n) \right] \\
 +  \int d^2r dt V({\bf r})\rho({\bf r})
\end{multline}
\end{widetext}
where the transverse (longitudinal) displacements $u_T$ ($u_L$)
are related to the cartesian displacement ${\bf u}$ as follows:
\begin{equation} \label{eq:decompdens}
 u_\alpha({\bf q}) =  u^L({\bf q})\hat{{\bf q}}_\alpha  + u^T({\bf q})
 \epsilon_{\alpha\beta}\hat{{\bf q}}_\beta
\end{equation}
$\hat{{\bf q}} = {\bf q}/|{\bf q}|$ is the unit vector along ${\bf
q}$ and $\epsilon_{\alpha\beta}$ is the fully antisymmetric tensor
with $\epsilon_{xy} = 1$. These two modes can be interpreted as
the shear ($u_T$) and the compression ($u_L$) modes respectively.
The Matsubara frequencies are $\omega_n = \frac{2\pi n}{\hbar
\beta}$ with $\beta=1/T$ being the inverse temperature. The terms
quadratic in the frequency represent the kinetic energy of these
modes. $\rho_m = {m \over {\pi a^2}},\rho_c = {e \over {\pi a^2}}$
are the mass and charge densities  respectively. The elastic
energies of these modes are given\cite{bonsall_elastic_wigner} by
\begin{equation}\label{moduli}
 \begin{split}
   \Omega_L({\bf q}) &= d_L |{\bf q}| + c_L {\bf q}^2\\
   \Omega_T({\bf q}) &= c_T {\bf q}^2
 \end{split}
\end{equation}
where  $c_L,c_T,d$ are elastic constants and, $\epsilon_0$
is the dielectric constant of the
substrate. For the classical crystal on the triangular lattice,
one has\cite{bonsall_elastic_wigner}  $c_L= - 0.18\frac{\rho_c^2
a}{\epsilon_0}  $, $c_T = 0.04 \frac{\rho_c^2a}{\epsilon_0}$ and
$d=\frac{\rho_c^2}{\epsilon_0}$.  These values for the elastic
modulii in (\ref{moduli}), are valid only for low densities, deep
in the crystal phase. For arbitrary densities, the elastic
constants depend on the scale $\xi$ and change drastically as one
approaches melting. This point should be taken into account while
comparing theoretical results for the  density dependence of
various quantities, with experiments. Unfortunately a rigorous
estimate of the elastic constants as a function of the density is
still lacking. The linear ${\bf q}$ dependence in the compression
mode $\Omega_L({\bf q})$ arises from the coulomb repulsion. This
is because a longitudinal deformation changes the density profile
of the crystal which costs Coulomb energy. On the contrary,
transverse modes can be excited without changing the density and
are thus purely elastic.

Finally, the last term in the action  describes the coupling to
disorder, modelled here by a random potential $V$. The density of
the particles $\rho({\bf r})$ is given by
\begin{equation}
 \rho({\bf r}) = \sum_i \overline{\delta}({\bf r} - {\bf R}_i -
 {\bf u}_i)
\end{equation}
where $\overline{\delta}$ is a $\delta$-like function of range
$r_f$. Since the disorder can vary at a lengthscale $r_f$  which
can \emph{a priori} be shorter or comparable to the lattice
spacing $a$, the continuum limit ${\bf u}({\bf R}_i) \to {\bf
u}({\bf r})$, valid in the elastic limit $|{\bf u}({\bf R}_i) -
{\bf u}({\bf R}_{i+1}) \ll a$, should be taken with care in the
disorder term\cite{giamarchi_vortex_short,giamarchi_vortex_long}.
This is done using the decomposition of the density in terms of
its Fourier components
\begin{equation} \label{eq:fourdens}
 \rho({\bf r})\simeq \rho_0 - \rho_0\nabla\cdot {\bf  u} + \rho_0
 \sum_{{\bf K} \neq 0} e^{i K\cdot({\bf r} - {\bf u}({\bf r}))}
\end{equation}
where $\rho_0$ is the average density and ${\bf K}$ are the
reciprocal lattice vectors  of the crystal.

The above formula shows that one should distinguish between various parts
of the disorder. Disorder that varies at lengthscales much larger
than the lattice spacing couples only to the $\rho_0\nabla\cdot u$
term. Although such a term can modify the structure of the
crystal quite strongly, it does not lead to pinning and disappears
from the transport properties. Pinning, and thus the dominant
contribution to transport coefficients, is however achieved by the
disorder that varies on lengthscales comparable to the lattice
spacing. In heterojunctions, such a disorder is expected to be
present due to interface roughness arising from epitaxial growth
of the semiconducting layers sandwiching the electron gas. The
real system is further complicated by the presence of a long range
disorder potential arising from the presence of ineffectively
screened dopants outside the plane of the electron gas. However
such a disorder varies slowly compared to the scale of the lattice
spacing, and we will thus assume that we can neglect it as far as
the transport properties are concerned. Moreover, if the disorder
is weak and leads to collective pinning one can assume a gaussian
distribution for $V$, since each volume of the system will average
over a large number of independent impurities. In this case,
\begin{equation}
 \overline{V({\bf r}) V({\bf r}^\prime)} = \Delta_{r_f}({\bf
 r}-{\bf r}^\prime)
\end{equation}
where $\Delta_{r_f}$ is effectively a delta function with a range
$r_f$.

\subsection{Gaussian variational method}

We treat (\ref{eq:ham}) using a variational
method\cite{giamarchi_columnar_variat,chitra_wigner_hall}. We
present here only the main steps of the treatment and refer the
reader to Ref.~\onlinecite{chitra_wigner_long} for details.  Many
of the technical details and subtleties of the method can be found
in the literature (see e.g.
Ref.~\onlinecite{giamarchi_book_young,giamarchi_quantum_pinning}
for a review). We first average (\ref{eq:ham}) over disorder by
introducing replicas. This averaging results in an effective
action which involves interactions between the $n$ replicas, given
by
\begin{widetext}
\begin{multline}\label{eq:seff}
 S = \frac12 \sum_{\omega_n} \int_{\bf q} \sum_a
 u_T^a({\bf q},\omega_n)[\rho_m\omega_n^2 +\Omega_T({\bf q})] u_T^a(-{\bf q},-\omega_n)  +
 u_L^a({\bf q},\omega_n)[\rho_m\omega_n^2 +\Omega_L({\bf q})] u_L^a(-{\bf q},-\omega_n)\\
  - \frac{\rho_0^2}{2} \int d^2r \int_0^{\beta }\int_0^{\beta} d\tau d\tau' \sum_{a,b, {\bf K}} \Delta_K \cos\left[{\bf
 K}\cdot\left(u^a({\bf r},\tau) - u^b({\bf r},\tau')\right) \right]
\end{multline}
\end{widetext}
The replica indices, $a,b$  run from $1$ to $n$.  The size of the
particles $\xi$ and the finite correlation length of the disorder
restrict the sum over $K$ to values of $K$ smaller than $K_{\rm
max} \sim \pi/\max(r_f,\xi)$. $\Delta_K$ is taken as a constant
$\Delta_K = \Delta$ for $K < K_{\rm max}$ and zero otherwise. The
physical disorder averages are recovered in the limit $n\to 0$.

We now search  for a variational solution to  (\ref{eq:seff}) by
using the best quadratic action approximating (\ref{eq:seff}). We
use the trial action
\begin{equation}\label{eq:trial}
 S_0= \frac{1}{2 \beta} \int_{{\bf  q}} \sum_{n,\mu,\nu} u_{\mu}^a
 ({\bf q},\omega_n) (G^{-1})^{ab}_{\mu \nu}({\bf q},\omega_n)
 u_{\nu}^b(-{\bf q}, -\omega_n)
\end{equation}
where the whole Green's function $(G^{-1})^{ab}_{\mu \nu}({\bf
q},\omega_n)$ are variational parameters. The variational free
energy is now given by
\begin{equation}\label{eq:free}
 F_{\rm var} = F_0 + \langle S-S_0 \rangle_{S_0}
\end{equation}
The variational parameters are then determined by the saddle point
equations
\begin{equation} \label{eq:saddle}
 \frac{\partial F_{\rm var}}{\partial (G^{-1})^{ab}_{\mu
 \nu}({\bf q},\omega_n)}=0
\end{equation}
The  pertinent solution of these saddle point equations
(\ref{eq:saddle}) solved in the limit of the number of replicas
$n\to 0$, breaks replica
symmetry\cite{giamarchi_columnar_variat,chitra_wigner_hall,chitra_wigner_long}.

The connected Green's functions  defined as $(G_c^{-1})_{\mu \nu}
= \sum_b (G^{-1})_{\mu\nu}^{ab}$ for the two modes are given by
\begin{equation}\label{gc}
\begin{split}
 G_{cT}({\bf q},i\omega_n) &= \frac{1}{  {\rho_m}\omega_n^2 + \Omega_T({\bf q})
 +I(i\omega_n) +\Sigma(1- \delta_{n,0})}  \\
 G_{cL}({\bf q},i\omega_n) &= \frac{1}{{\rho_m}\omega_n^2
 +\Omega_L({\bf q})+I(i\omega_n) +\Sigma(1- \delta_{n,0})}
\end{split}
\end{equation}
where the disorder induced pseudogap is found to be
\begin{equation}\label{sigmab-0}
 \Sigma = c_T R_{c}^{{-2}}= c_T (2\pi^2)^{-{1 \over 6}} R_a^{-2} (a/\xi_0)^6
\end{equation}
with $\xi_0= \max[r_f,\xi]$, $c_T\simeq \frac{\rho_c^2a}{\epsilon_0}
$ and $R_a \simeq c_T/\pi n^2 \sqrt\Delta $. As explained in
Refs.~\onlinecite{giamarchi_columnar_variat,chitra_wigner_long},
the function  $I(i\omega_n)$  is determined by the following
self-consistent equation  derived in the semi-classical limit:
\begin{widetext}
\begin{multline}\label{iwn1}
 I(i\omega_n) = 2 \pi c_T \Sigma \int_{\bf q} \left[{1 \over
 {\Omega_T({\bf q}) + \Sigma}} + {1 \over { \Omega_L({\bf q})+ \Sigma}} \right.  \\
  \left. - \frac{1}{\Omega_T({\bf q})  +
 \rho_m\omega_n^2 +I(i\omega_n) + \Sigma} -\frac{1}{\Omega_L({\bf q})+
 \rho_m \omega_n^2 +I(i\omega_n) + \Sigma} \right]
\end{multline}
\end{widetext}
In the limit of small $\omega_n$, the above equation can be solved analytically to obtain
\begin{equation}\label{iw-0}
 I(i\omega_n)= \sqrt{2\rho_m \Sigma[1 + \frac{ 4c_T\Sigma}{d^2 }\log(\frac{d^2}{c_T \Sigma})]} \vert \omega_n \vert
\end{equation}
This equation for $I$ can be continued to real frequencies and
then solved numerically.
The solution of the variational equations as obtained above thus
allows us to extract numerous  physical quantities of the disordered
system. We will first discuss the compressibility in the forthcoming
section and the  transport will be discussed in
\sref{sec:transport}.

\section{Compressibility}\label{sec:comp}

In this section we discuss the compressibility of the pinned Wigner
crystal. Although the compressibility is usually simply related to
density-density correlations, here one should take special care in
defining this quantity due to two complications inherent to the
Wigner crystal: (i) for charged systems it is crucial to know whether
one keeps the system neutral or not when letting the density
fluctuate. (ii) since in presence of disorder one expects glassy
properties, the question of the timescale  over  which the
measurements are performed is relevant.

The naive way to define the compressibility $\kappa$ would be to
add to the Hamiltonian the standard chemical potential term
\begin{equation}
 H = - \mu \int d^2r (\rho({\bf r}) -\rho_0)
\end{equation}
and to compute the average change in density $\langle
\rho({\bf r})-\rho_0 \rangle$. In linear response, the compressibility is
given by the density density correlator. Using
(\ref{eq:decompdens}) and (\ref{eq:fourdens}) the compressibility
is directly related to the correlation of the longitudinal
displacements. For the pure system this is easily computed to be
\begin{equation}
 \kappa \propto \lim_{q\to 0, \omega_n = 0} \frac{q^2}{\omega_n^2 + \Omega_L({\bf q})}
\end{equation}
Since  $\Omega_L({\bf q}) \propto q$, the compressibility is zero.
This arises from the fact that the change of density induced by
$\mu$ occurs for a constant neutralizing background. The combined system
of the  crystal  plus the background thus becomes charged which costs too much energy and inhibits
this kind of density fluctuation.

\begin{figure}
 \centerline{\includegraphics[width=\smallwidth]{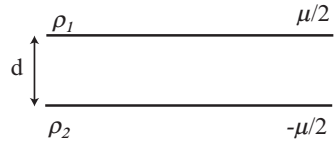}}
 \caption{Capacitance measurement, which gives access to the
 compressibility of the system. A voltage difference $\mu$ is
 applied to a capacitor. Here for simplicity, the capacitor is made
 of two planes of the 2DEG.} \label{fig:capacitance}
\end{figure}

To have a non-zero compressibility, even for the pure system, it
is thus important to change the density of the Wigner crystal,
while at the same time changing the background to maintain
neutrality. An experimental way of doing this is to perform a
capacitance measurement\cite{eisenstein_hall_compressibility}. We
show here how to compute such a capacitance for the case of two
Wigner crystals planes\cite{giamarchi_wigner_review} separated by
a distance $d$, as shown in \fref{fig:capacitance}.
Though the  experimental setup (two layers and a gate) considered
in Ref.~\onlinecite{eisenstein_hall_compressibility} is slightly
different from the one considered here, we expect the results
obtained here to hold for this experimental geometry as well,
since the crucial ingredient, namely ensuring the neutrality of
the system is preserved.

The Hamiltonian of the system depicted in \fref{fig:capacitance}
is thus
\begin{widetext}
\begin{equation} \label{eq:hamdep}
 H = H_1^0 + H_2^0 +  \frac12\sum_{(\alpha,\beta)=1,2} \int_{{\bf r},{\bf r}'} V_{\alpha \beta }({\bf r}-{\bf r}')
 [\rho_\alpha({\bf r}) - \rho_0][\rho_\beta({\bf r}') - \rho_0]
 + \frac{\mu}2 \int_{\bf r} [\rho_1({\bf r}) - \rho_2({\bf r})]
\end{equation}
\end{widetext}
where $H^0_{1,2}$ are the elastic Hamiltonians of each 2DEG  excluding
the Coulomb interaction. $V_{\alpha \beta }$ is the usual Coulomb interaction in and between the planes. If one assumes that the system is neutral
in the absence of $\mu$, then  within linear response, the charge on one plane when a
potential $\mu$ is applied is
\begin{equation} \label{eq:congen}
 \langle \rho_1 \rangle = \frac{\mu}2 [ \langle \rho_1 \rho_1 \rangle -
 \langle \rho_1 \rho_2 \rangle ].
\end{equation}
Thus (\ref{eq:congen}) gives directly the capacitance $C = \langle
\rho_1 \rangle/\mu$. More generally, we define the capacitance for
finite $q$ and $\omega$ (e.g. corresponding to a time and space
dependent potential $\mu$)
\begin{equation} \label{eq:capadef}
 C(\omega,{\bf q}) = \frac12[\langle \rho_1 \rho_1 \rangle_{\omega,{\bf q}} -
 \langle \rho_1 \rho_2 \rangle_{\omega,{\bf q}}]
\end{equation}
where the subscripts ${\omega,{\bf q}}$ indicate that the
correlation functions are evaluated for finite momentum and
frequency. For a metal, a RPA evaluation of (\ref{eq:capadef})
gives back\cite{giamarchi_wigner_review} the standard formula for
the compressibility of a charged system. We recall the calculation
in \aref{ap:rpacoul} for convenience. One can use this general
formula to compute the capacitance for the Wigner crystal. Since
one is interested in the limit of small $q$ and $\omega$ one can
use the decomposition of the density (\ref{eq:fourdens}). The
Hamiltonian (\ref{eq:ham}) becomes
\begin{multline} \label{eq:hamplanes}
 H = H_1^{sr} + H_2^{sr} +  \\ \frac{\rho_0^2}2\sum_{(\alpha,\beta)=1,2} \int_{{\bf r},{\bf r}'} V_{\alpha \beta }({\bf r}-{\bf r}')
 [\nabla \cdot {\bf u}_\alpha({\bf r})][\nabla \cdot{\bf  u}_\beta({\bf r}')]
\end{multline}
where $H_{1,2}^{sr}$ is the part of the Hamiltonian that does not
contain the \emph{long range} (i.e. for ${\bf q }\sim 0$)) of the
Coulomb interaction. ${\bf u}_1$ and ${\bf u}_2$ denote the displacement
vectors in the two planes. Since $V_{11}({\bf q}) = V_{22}({\bf q})
\propto 1/|{\bf q}|$, the third term in (\ref{eq:hamplanes})
obviously  yields the part proportional to ${\bf q}$ in the  bulk
modulus  for an isolated plane (see (\ref{moduli})). Note that in
writing (\ref{eq:hamplanes}) we have neglected the short range
part of the interplane interaction
\begin{equation}
 H_{1,2}^{SR} = \int_{{\bf r},{\bf r}'} V_{\alpha \beta }({\bf r}-{\bf r}') \rho_0^2 \sum_{{\bf K},{\bf K}'\ne 0}
 e^{i {\bf K} \cdot ({\bf r} - {\bf u}_1({\bf r}))} e^{-i {\bf K}' \cdot ({\bf r}' - {\bf u}_2({\bf r}'))}
\end{equation}
The translational invariance restricts the sum to ${\bf K} = {\bf
K}'$ terms. This essentially  contains the Fourier transform of
the interplane potential which using (\ref{eq:cancelcoul}) is
found to behave as
\begin{equation}
 \int_{\bf r} V_{12}({\bf r}) e^{i {\bf K} \cdot  {\bf r}} = \frac{(2\pi) e^{-K d}}{K}
\end{equation}
Thus if the planes are at a distance $d$ much larger than the
lattice spacing $a$ of the WC, this term is obviously much smaller
than the long range part of the intraplane interaction, and can
 be neglected. It can also be neglected if one of the planes
is an homogeneous electron gas (i.e. a simple metal). In the case
where the two planes are close enough this term should be retained
and can lead to interesting effects such as the locking of the two
Wigner crystals together\cite{falko_bilayer}. The in-plane
coupling of the higher harmonics of the density contribute to
$H^{SR}$ and generate  the regular non-singular part of the
hamiltonian (i.e. the part proportional to $q^2$ in the elastic
coefficients). In fact this approach is synonymous with the method
used in Ref.~\onlinecite{bonsall_elastic_wigner} to calculate the
elastic coefficients. The derivation is done in \aref{ap:bonsder}.

Let us now use the general formula (\ref{eq:capadef}) to compute
the capacitance of   the pure system. From
(\ref{eq:fourdens}), the long wavelength part of the density is
given by
\begin{equation} \label{eq:denssmallq}
 \rho_\alpha ({\bf q}) = - i \rho_0  q u_{\alpha L}({\bf q})
\end{equation}
One thus needs only the longitudinal part of the action to calculate the
compressibility.
\begin{widetext}
\begin{equation}
 S =  \left(\begin{array}{cc} u_L^1({\bf q}) & u_L^2({\bf q}) \end{array}\right)
 \left(\begin{array}{cc} \omega_n ^2 + \csr({\bf q}) +  \rho_0^2 q^2 V_{11}({\bf q})& \rho_0^2 q^2 V_{12}({\bf q}) \\
                       \rho_0^2 q^2 V_{12}({\bf q})   & \omega_n^2 + \csr({\bf q}) +  \rho_0^2 q^2 V_{11}({\bf q})
       \end{array} \right)
 \left(\begin{array}{c} u_L^1(-{\bf q}) \\ u_L^2(-{\bf q}) \end{array} \right)
\end{equation}
\end{widetext}
where $\csr({\bf q})$ is the ``short range'' part of the elastic
coefficients (\ref{moduli}). Using (\ref{eq:denssmallq}) and
(\ref{eq:capadef}) one obtains the capacitance
\begin{equation}
 C(i \omega_n,{\bf q}) = \frac12 \frac{\rho_0^2 q^2}{\rho_m \omega_n^2 + \csr({\bf q}) +  \rho_0^2 q^2 [V_{11}({\bf q}) - V_{12}({\bf q})]}
\end{equation}
The thermodynamic compressibility is given by the value of $C$  for
$\omega_n = 0$. Note that in this case one obtains the same value
by considering  the retarded correlation function (doing the
analytic continuation $i\omega_n \to \omega + i \delta$) and
taking the limit $\omega \to 0$ first (for a fixed $q$). The
divergence arising from the long range part of the Coulomb
potential now cancels since
\begin{multline} \label{eq:cancelcoul}
 V_{11}({\bf q}) - V_{12}({\bf q}) = \int d^2r e^{i {\bf q}  \cdot {\bf r}} [\frac1r -
 \frac1{\sqrt{r^2+d^2}}] = \\ \frac{(2\pi)(1 - e^{-qd})}{q}
\end{multline}
is non divergent when $q \to 0$.
This of course traduces the fact that the global system remains
neutral when the potential is applied. Using (\ref{eq:cancelcoul})
one obtains
\begin{equation}
 C(\omega_n = 0,q \to 0) \to \frac{\rho_0^2 q^2}{2\csr({\bf q}) + 4\pi d (\rho_0^2 q^2)}
\end{equation}
One thus recovers that the inverse capacitance is the sum of a
purely geometric term and an electronic one (see
(\ref{eq:geom})). The electronic capacitance is simply given by
\begin{equation} \label{eq:capaelec}
 C_{el} = \lim_{q\to 0} \frac{\rho_0^2 q^2}{2 \csr(q)} = \frac{\rho_0^2}{2 c_L} \propto - \frac{\epsilon_0}{e^2 a }
\end{equation}
This term, which for a normal metal is simply related to the
static compressibility, is thus negative for a Wigner crystal, if
one uses the classical estimates\cite{bonsall_elastic_wigner} for
the elastic coefficients. The fact that a system of discrete
charges can lead to such effects has been noted before for
classical Wigner crystals (see e.g.
Ref.~\onlinecite{nguyen_wigner_biology_houches} and references
therein).

Let us now turn to the disordered case. In the presence of
disorder, the same calculation can be repeated to obtain the
compressibility. To do this, we assume that the disorder
potentials in each layer is drawn from independent distributions
 and  repeat the variational calculation of the previous
section. The trial action (\ref{eq:trial}) has now two additional
indices that denote the two layers. For simplicity, we will only
denote these indices here, all the others being implicit. The
trial action is thus (in these indices) a two by two matrix with
the Green's functions $G^{-1}_{\mu\nu}$ where $\mu,\nu = 1,2$.
Obviously $G_{11}=G_{22}$ and $G_{12}=G_{21}$. From
(\ref{eq:capadef}) and the inversion of the two by two matrix the
capacitance is thus given by
\begin{equation} \label{eq:compvar}
 C(\omega,{\bf q}) = \frac{\rho_0^2 { q}^2}2[G_{11}-G_{12}]
 = \frac{\rho_0^2 { q}^2}2[\frac1{G^{-1}_{11} - G^{-1}_{12}}]
\end{equation}
where all other indices (replica, longitudinal, $q$, $\omega$) are
implicit. The variational procedure  can now be repeated to
determine $G_{11}$ and $G_{12}$. Because the disorder is
independent from plane to plane,  the trial free energy  has the following matrix
structure
\begin{widetext}
\begin{equation}
 \left(\begin{array}{cc}
  [\rho_m \omega_n ^2 + \csr({\bf q}) +  \rho_0^2 q^2 V_{11}({\bf q})] G_{11} + F[G_{11}]
  & \rho_0^2 q^2 V_{12}({\bf q}) G_{12} \\
  \rho_0^2 q^2 V_{12}({\bf q}) G_{21} & [\rho_m \omega_n ^2 + \csr(({\bf q}) +  \rho_0^2 q^2 V_{11}({\bf q})]
  G_{22} + F[G_{22}]
 \end{array}\right)
\end{equation}
\end{widetext}
where $F$  is the same  as the term induced by disorder averaging
in a single plane. The minimization (\ref{eq:saddle}) now gives
the self consistent equations
\begin{equation} \label{eq:solvarcomp}
\begin{split}
 G^{-1}_{11} & = [\rho_m \omega_n ^2 + \csr({\bf q}) +  \rho_0^2 { q}^2
 V_{11}({\bf q})] + F'[G_{11}] \\
 G^{-1}_{12} & = \rho_0^2 { q}^2 V_{12}({\bf q})
\end{split}
\end{equation}
Note that the in-plane inverse Green's function is  identical to the one in the
absence of interplane interactions, and thus is given by the
variational solution of the previous section. The interplane inverse Green's
function is trivial and simply given by the long range part of the
interaction between the planes.

In the presence of disorder, the compressibility is given by the
connected Green's function\cite{giamarchi_quantum_pinning}. Using
(\ref{eq:compvar}), (\ref{eq:solvarcomp}) and (\ref{gc}) one
obtains
\begin{widetext}
\begin{equation}
 C(\omega_n,{\bf q}) = \frac12 \frac{\rho_0^2 { q}^2}{\rho_m \omega_n^2 +
 \csr({\bf q}) + \rho_0^2 { q}^2 [V_{11}({\bf q}) - V_{12}({\bf q})] + I(i\omega_n)
 +\Sigma(1- \delta_{n,0})}
\end{equation}
\end{widetext}
As for the pure system, we see that the geometric capacitance contributes to the total capacitance. Isolating  the electronic part of the
capacitance we find
\begin{equation} \label{eq:capaelecdis}
 C_{el}(\omega_n,{\bf q}) = \frac{\rho_0^2 { q}^2/2}{\csr({\bf q}) + I(i\omega_n)
 +\Sigma(1- \delta_{n,0})}
\end{equation}
(\ref{eq:capaelecdis}) thus shows that in the presence of disorder
one has to distinguish between the thermodynamic capacitance and
the dynamical one. If one considers the thermodynamic capacitance,
then (\ref{eq:capaelecdis}) should be computed for $\omega_n = 0$.
In that case all contributions from the disorder disappear in
(\ref{eq:capaelecdis}) and the capacitance is identical, within
the variational approximation, to that of the pure system. On the
other hand, if one considers a capacitance measurement in response
to a modulation at finite frequency, which is certainly the case
experimentally, one has to take the analytic continuation $i
\omega_n \to \omega + i \delta$ first. In this case one obtains
\begin{equation} \label{eq:capaelecdisreal}
 C_{el}(\omega,{\bf q}) = \frac{\rho_0^2 { q}^2/2}{\csr({\bf q}) + I(\omega)
 +\Sigma}
\end{equation}
As for the pure case, one should  take the limit $\omega \to 0$
first. In this limit, the function $I(\omega)$ is regular and
tends to zero. Taking  $q \to 0$, we see that since the  mass term
$\Sigma$  prevails  in the denominator  the compressibility is
zero. This is a consequence of the fact that the pinning by
disorder renders the system inflexible to charge modulations. Note
that contrary to the pure case, the thermodynamic response and the
slow dynamic one are not equivalent in the disordered case. This
is not surprising considering the glassy nature of the system.

As for the effect of a magnetic field on the compressibility,
naively one would expect that it has no effect since the
compressibility is a static property. This is what is found within
the variational approximation. However it is important to note
that, although there is no explicit magnetic field dependence of
the compressibility, there is still a variation of the thermodynamic
compressibility with the  magnetic field  accruing from the field dependence
of the elastic constants of the crystal.

\section{Transport}\label{sec:transport}

We now focus on the consequences of disorder for the zero
temperature transport properties.  This is of particular
importance to the materials which exhibit the zero field
metal-insulator transition where transport is indeed the main
probe of the physics. Up to now, the experimental emphasis has
been on finite temperature resistivity and magneto-resistance
measurements. However for the Wigner crystal such quantities are
difficult to compute theoretically since they are dominated by the
defects in the system\cite{chitra_wigner_long}. It is thus
difficult to use them as a probe of the Wigner crystal nature of
the underlying phase. As explained in
Ref.~\onlinecite{chitra_wigner_long,giamarchi_varenna_wigner_review}
the optical conductivity does not suffer from such a problem and
can be reliably computed theoretically from the elastic
Hamiltonian of the crystal. Here, we present results for the
dynamical conductivity, surface acoustic wave measurements and the
power radiated by the crystal.  These measurements  done in the
quantum Hall samples at high fields were instrumental in
clarifying  the physics of these systems
\cite{li_optical_wigner,li_conductivity_wigner_magneticfield,li_conductivity_wigner_density,andrei_saw,williams_wigner_threshold}.
Clearly similar measurements are called for in  the zero field
samples, and would  be crucial to understand the physics of the
insulating phase.

\subsection{Conductivity}

\begin{figure}
 \centerline{\includegraphics[angle=-90,width=\normwidth]{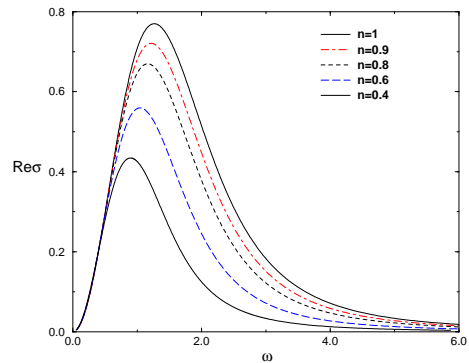}}
 \caption{The real part of the conductivity $\sigma(\omega)$ (in  units of $e^2/\sqrt{A_2 m}$ ) as a function of the frequency
 measured in units of $\sqrt{\frac{A_2}{m}}$  for the regime where the disorder correlation length $r_f < \xi$
 for various electronic densities $n$(in units of $(\pi a_B^2)^{-1}$). These curves have been obtained using the classical values of the elastic moduli and the
   parameter $A_2$ is given in the text.} \label{fig:var}
\end{figure}

\begin{figure}
 \centerline{\includegraphics[angle=0,width=\normwidth]{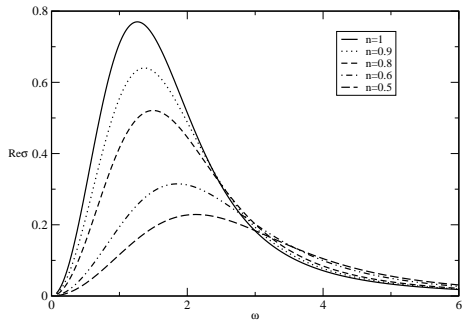}}
 \caption{ The real part of the conductivity $\sigma(\omega)$ (in  units of $e^2/\sqrt{A_1 m}$ ) as a function of the frequency
 measured in units of $\sqrt{\frac{A_1}{m}}$  for the regime where
 the disorder correlation length $r_f > \xi$ for various densities $n$ ( in units of $(\pi a_B^2)^{-1}$). As before, these curves have been obtained using the classical values of the elastic moduli and the
   parameter $A_1$ is given in the text.} \label{fig:var1}
\end{figure}

The conductivity  of the disordered crystal can be obtained from
the displacement-displacement correlation
function\cite{chitra_wigner_long} and is given by:
\begin{equation}
 \sigma_{\alpha \beta}(\omega)= i \rho_c^2\omega G_{\alpha \beta} (q=0,
 \omega+ i\epsilon)
\end{equation}
where $\mu\nu= x,y$ and $G_{\mu\nu}({\bf q },\omega)= \langle
u_\mu({\bf q },\omega)u_\nu({\bf q },\omega)\rangle$ are the
displacement Green's function. Since, there is no magnetic field,
the longitudinal resistivities $\sigma_{xx} (\omega) = \sigma_{yy}
(\omega) = \sigma (\omega)$. These functions are related to the
connected Green's functions of (\ref{gc}) and using an analytic
continuation $i\omega_n \to \omega +i \epsilon$, we obtain
\begin{equation} \label{eq:conduct}
 \sigma(\omega)=i\rho_c^2\frac{\omega}  {- \rho_m\omega^2 + \Sigma
 + I (\omega)}
\end{equation}
As seen in other disordered elastic
systems\cite{giamarchi_columnar_variat,chitra_wigner_long}, the
conductivity is completely determined by $\Sigma$ and $I(\omega)$.
For the pure crystal, $\Sigma = I(\omega_n) = 0$, and one recovers
that the real part of the conductivity exhibits a $\omega=0$ Drude
peak and zero finite frequency conductivity, while the imaginary
part varies as $1/\omega$. In the presence of disorder, the
crystal is pinned collectively forbidding any sliding motion of
the crystal. Consequently, the Drude peak is annihilated and the
dc conductivity ${\rm Re}\sigma(\omega=0)=0$. This results in a
non-zero finite frequency conductivity and  the appearance of a
peak at a new scale $\omega_p \simeq \sqrt{\frac{\Sigma}
{\rho_m}}$ called the pinning frequency\cite{fukuyama_pinning}.
The pinning frequency is thus determined by (\ref{sigmab-0}).  The
pinning peak is broadened by the dissipative term  $I(\omega)$
which also generates, as can be seen from (\ref{eq:conduct}), a
slight shift of the pinning peak from $\omega_p$ due to the real
part of $I(\omega)$. Using (\ref{iw-0}), we find that for  $\omega
\ll \omega_p$.
\begin{equation}
\begin{split}
 {\rm Re}\sigma (\omega) &\sim \rho_c^2
 \sqrt{2\rho_m \Sigma[1 + \frac{4c_T \Sigma}
 {d^2 }\log(\frac{d^2}{c_T \Sigma})]} {{\omega^2}
 \over {\Sigma^2}} \\
 {\rm Im}\sigma(\omega) &=
 \rho_c^2 {\omega \over \Sigma} \\
\end{split}
\end{equation}
while for high frequencies $\omega \gg \omega_p$
\begin{equation} \label{cxx1}
 {\rm Re}\sigma(\omega)  \sim {\rho_c^2 \over \rho_m^2\omega^3}
\end{equation}
The behavior of the conductivity for all intermediate values of
$\omega$ can be obtained by solving numerically (\ref{iwn1}) and
using the solution for $I(\omega)$ in (\ref{eq:conduct}).
Referring to (\ref{sigmab-0}), we note that the result for
$\Sigma$ depends on whether the disorder correlation length is
larger or smaller than the size of the particles of the crystal
($r_f \lessgtr \xi$). This has serious implications for the
density dependence of the conductivity as can be inferred from
Figs.~\ref{fig:var} and ~\ref{fig:var1} where, we plot the
conductivity for different values of the density in the two
regimes $r_f < \xi$
and $r_f > \xi$.

The density dependence of the pinning frequency and the height of
the pinning peak can now be evaluated in a straightforward manner.
The peak height is given by the expression
\begin{equation}\label{eq:peakheight}
 P = \sigma(\omega_{p}) = \rho_{c}^2 \sqrt{\frac{\Sigma}{\rho_m}}
 \rm {Im}\frac{1}{I(\omega_p)}
\end{equation}
If we neglect corrections to the pinning frequency arising from $I$, and assume, then $I(\omega_p)= \Sigma Z$ where $Z$ is some complex number,  and in this case,
the leading density dependence of the peak height is
given by
\begin{equation}
 P \propto \frac{\rho_{c}^2} {\sqrt{\rho_m \Sigma}}
\end{equation}
The behavior of the peak height is thus entirely determined by
$\Sigma$. To obtain the actual density dependence, one needs to
know the elastic moduli of the crystal. Using the results of
Ref.~\onlinecite{bonsall_elastic_wigner}, we find that in the
regime, $r_f > \xi$, the electrons see the bare disorder and
\begin{equation}\label{eq:1}
 \begin{split}
 \Sigma &= A_1 n^{{-\frac12}}  \\
 R_{c} &= \sqrt{\frac{b e^2}{A_1\epsilon \sqrt{\pi}} n}
\end{split}
\end{equation}
where $A_1= (2\pi^2)^{-\frac16} \Delta \epsilon/ b\sqrt{\pi}e^2
r_f^6$ and $b=0.04$ for the classical triangular lattice. This leads to
\begin{equation}
\begin{split}
 \omega_{p} & \propto   n^{{-\frac{3}{4}}}  \\
 P & \propto  n^{\frac{7}{4}}
\end{split}
\end{equation}
Consequently, the peak position shifts to higher frequencies  and
the peak height decreases with  decreasing density as is seen in
\fref{fig:var1}.

In the opposite regime where the disorder correlation length
$r_{f}< \xi$, the effective particle size being larger, the
particle sees an averaged effective disorder. Using the classical
values for the elastic moduli, we find that
\begin{equation}\label{eq:2}
\begin{split}
 \Sigma &= A_2  n^{\frac{7}4}  \\
 R_{c} &=\sqrt{\frac{b e^2}{A_2\epsilon \sqrt{\pi}}} n^ {-\frac{1}8}
\end{split}
\end{equation}
where $A_2= (2\pi^2)^{-\frac16} \Delta
\epsilon\pi^\frac74 (C/a_B)^\frac32 / b e^2 $ and $C=1.88$ for the triangular lattice cf. \ref{en-lowd}.  Hence
\begin{equation}
\begin{split}
 \omega_{p} &\propto  n^{{\frac{3}{8}}} \\
 P &\propto  n^{\frac58}
\end{split}
\end{equation}
This implies that as density is decreased, the pinning peak shifts
to lower frequencies accompanied by a concomitant decrease of peak
height as shown in \fref{fig:var}. The density dependence of the
pinning peak is a test that can prove whether the insulating phase
is indeed a Wigner crystal.

Another interesting quantity is the threshold electric field for
depinning of the crystal. This field is
related\cite{larkin_ovchinnikov_pinning} to the parameters of the
problem by the relation  $E_{T}= c_T R_{c}^{{-2}}\xi$ where $\xi_0
= \max[r_f,\xi]$. Using (\ref{eq:1},\ref{eq:2}), we obtain
\begin{equation}
\begin{split}
 E_{T} &\propto  n^{{-\frac{1}{2}}} \quad,\quad r_{f} > \xi  \\
 E_{T} &\propto  n^{{\frac{11}{8}}} \quad,\quad r_{f} < \xi
\end{split}
\end{equation}
Note that for  $r_{f} > \xi$, the threshold field increases with
inverse density, implying that the crystal gets more and more
pinned in the low density regime. However, since $\xi$ increases
when the density decreases according to (\ref{size}) on has to
crossover to the other regime $r_f < \xi$. Depending on the
correlation length of the disorder, the system can thus cross over
from one kind of dynamical behavior to another as the density is
varied. In the first case, the pinning frequency and the threshold
field increase with decreasing density. When the particle size
exceeds the length scale of disorder, the pinning frequency and
the threshold decrease with decreasing density.  In both cases,
the peak height decreases with decreasing density. This is
compatible with the reduction of spectral weight with decreasing
density. Such a crossover would lead to a maximum in the threshold
field as the density is decreased.

If we now turn on a magnetic field perpendicular to the plane of
the crystal, the result for $\Sigma$ is unchanged as long as the
magnetic length $l_{c} > \xi$. For weak fields, the peaks in the
diagonal conductivity now occur at the frequencies
$\omega_{p}^{B}= \omega_{c} \pm \omega_{p}^{{B=0}}\pm
\frac{\omega_{c}^{2}}{8\omega_{p}^{B=0}}$. For strong enough
fields such that $l_{c}\leq \xi$, we revert back to the usual
strong field physics discussed in
Ref.~\onlinecite{chitra_wigner_long}.

\subsection{Surface Acoustic Wave Measurements}\label{sec:saw}

\begin{figure}
 \centerline{\includegraphics[width=\normwidth]{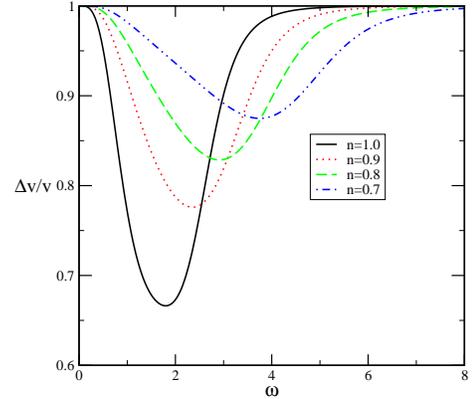}}
 \caption{The typical profile of the  SAW velocity shift (in arbitrary units calculated for $\sigma_m=1$ and
 $\alpha^2 =2$  ) as a function
 of the frequency $\omega = v q$ and density. In both the  regimes  of disorder, the position of the
 minimum shifts to higher frequencies and the depth decreases as density is
 decreased. The quantitative behavior will be governed by the value of $\Sigma$.
 correlation length $r_f > \xi$. The curves are plotted for
 various values of the density $n$.}  \label{fig:saw1}
\end{figure}

Another physical property that can be extracted from our results
pertains to surface acoustic waves (SAW). SAW measurements were
used in the past to obtain the crystal dispersion
relations\cite{andrei_saw}. In SAW measurements,  the system is
excited by a finite $q$ SAW with a frequency $\omega= v q/2 \pi$,
where $v$ is the velocity of the SAW. As the excitation traverses
the system, the SAW propagation is completely affected by the
piezoelectric interaction  resulting in an attenuation and a shift
in the SAW velocity. This shift $\delta v$ and the attenuation
rate $\kappa$ are  essentially  determined by the density response
function of the system and are given by\cite{halperin_lee_saw}
\begin{equation}\label{velshift}
\begin{split}
 \frac{\Delta v}{v} &= \frac{\alpha^{2}}{2} \frac{1}{1+
 [\frac{\sigma(q)}{\sigma_{m}}]^{2}} \\
 \kappa &= \frac{q\alpha^{2}}{2}
 \frac{\sigma(q)\sigma_{m}}{\sigma_{m}^{2}+
 \sigma(q)^{2}}
\end{split}
\end{equation}
where $\alpha$ is the piezoelectric coupling constant which
determines the interaction between the SAW and the electron system
and $\sigma_{m}= v (\epsilon+\epsilon_{0})$ where $\epsilon$ and
$\epsilon_{0}$ are the dielectric constants of the medium and
vacuum respectively. and  the momentum dependent longitudinal
conductivity of the crystal
 $\sigma(q)\equiv{\rm{Re}} \sigma(q,\omega= v q)$.
Notice that $\Delta v$ and $\kappa$
provide an indirect measurement of the conductivity at finite $q$.
Using the results of the preceding sections, we obtain
\begin{equation}\label{}
 \sigma (q,\omega) = \frac{i \rho_c^2 \omega q_{x}^{2}}{q^{2}} G_{L}
 (q,\omega) + \frac{i \rho_c^2 \omega q_{y}^{2}}{q^{2}} G_{T} (q,\omega)
\end{equation}
where
\begin{equation}
\begin{split}
 G_{L} (q,\omega) &= \frac{1}{\Sigma + \Omega_L(q) - \rho_{m}
 \omega^{2} + I(\omega)} \\
 G_{T} (q,\omega) &= \frac{1}{\Sigma + \Omega_T (q)- \rho_{m} \omega^{2} +I(\omega)}
\end{split}
\end{equation}
Using $I(\omega) = I_1(\omega) + i I_2(\omega)$ one gets
\begin{widetext}
\begin{equation}
 {\rm Re} \sigma(q,\omega=v q) = \frac{\rho_c^2 v }{q}
 \left[\frac{q_{x}^{2}I_{2} (vq)}{[\Sigma + q^{2} (c_L -\rho_{m}
 v^{2}) +dq +I_{1} (vq)]^{2}+ I_{2}(vq)^{2}} + \frac{q_{y}^{2}I_{2}
 (vq)}{[\Sigma + q^{2} (c_T -\rho_{m} v^{2}) +I_{1} (vq)]^{2}+
 I_{2}(vq)^{2}}\right]
\end{equation}
\end{widetext}
where $I_1$ and $I_2$ are the real and imaginary parts of
$I(\omega)$. For small frequencies since $I_{2}(\omega) \propto
\omega$, and $\Sigma$ is the dominant scale,
\begin{equation}
 {\rm Re}\sigma(q,\omega=v q) = \Gamma \frac{ v^{2} q^{2}}{\Sigma^{2}}
\end{equation}
where $\Gamma=\sqrt{2\rho_m \Sigma[1 + \frac{ 4c_T\Sigma}{d^2
}\log(\frac{d^2}{c_T \Sigma})]}$ is a density dependent parameter
independent of $q$. Consequently, one has
\begin{equation}\label{velshift1}
\begin{split}
 \frac{\Delta v}{v} &= \frac{\alpha^{2}}{2} \frac{1}{1+
 [\frac{\Gamma v^{2} q^{2}}{\Sigma^{2}\sigma_{m}}]^{2}} \\
 \kappa &= \frac{\alpha^{2}}{2}
 \frac{\Gamma  v^{2}q^{3}\Sigma^{2}\sigma_{m}}{ \Sigma^{4}\sigma_{m}^{2}+
 \Gamma^{2}v^{4}q^{4}}
\end{split}
\end{equation}
Using our results for $I(\omega)$ and the wave vector dependent
conductivity, we can thus calculate the shift and attenuation for
all values of $q$. The results are shown in \fref{fig:saw1}.
 In both the regimes $r_f \lessgtr \xi$,
despite a reduction in the the magnitude of the SAW anomaly with
decreasing density, the width of anomaly is  expected to be
strongly density dependent and is essentially dictated by the
competition of the disorder scale $r_f$ and the effective particle
size. These results provide an impetus for detailed SAW
experiments in the 2d electron gas systems. We reiterate that the
above derivation has been done with the classical values of the
elastic constants, and is thus valid deep in the crystal phase. A
full quantitative prediction would require a precise knowledge of
the variation of the elastic constants with density, a piece of
information missing at the moment.

\subsection{Power radiated by the 2d crystal} \label{sec:power}

Another possible probe of the physics of the electron gas concerns
thermometry. This involves a study of transfer of heat generated
by a current in  the electron gas to its three dimensional
environment. This occurs via phonon emission at any given
temperature and is related to the internal dynamics of the
electron gas. These measurements were done on the electron gas
both in the absence and presence of magnetic
fields\cite{chow_girvin_phonon,chow_girvin_exp,appleyard_phonon}.
The expression for the power radiated by the 2d gas at a
temperature $T$ into  cooler environment, considered to be at zero
temperature is given by \cite{chow_girvin_exp}:
\begin{equation}\label{power}
 P = \sum_{\bf Q}\omega({\bf Q})\vert M ({\bf Q})\vert^{2} n_{B} (\beta \hbar
 \omega({\bf Q})) G (q,\omega ({\bf Q}))
\end{equation}
where $G(q,\omega) \propto  \omega {\rm Re}(1/\sigma(q,\omega))$,
$q$ is the projection of the phonon momentum ${\bf Q}$ onto the
plane of the WC, $M$ is the electron phonon matrix element and
$n_{B}$ is the usual Bose distribution function. Following
Ref.~\onlinecite{chow_girvin_phonon}, we assume that the electron
phonon coupling is dominated by the piezoelectric effect and
$\vert M ({\bf Q})\vert^{2}\sim 1/Q$. In the absence of any
external field and for high mobility systems, $G \propto \omega^2$
for $\omega \equiv \omega({\bf Q}) \propto Q$ and and the
predicted power radiated $P\propto T^5$ was found to be in accord
with experiments. Based on power counting arguments, it was also
shown that the radiated power in the IQHE phase $P \propto T^4$.
In the light of these experiments, an obvious question is what is
the power radiated in the pinned crystal. A naive approach would
be to suppose that finite temperature contributions to  ${\rm Re}
1/\sigma$ are small at low temperatures, hence  permitting one to
approximate $G (q,\omega)$ by its zero temperature value. In this
case,   we find that  for the frequency range $0 < \omega <
\omega_p$, ${\rm Re} 1/\sigma(q,\omega(q))$ does not vary
significantly. Following the lines of
Ref.~\onlinecite{chow_girvin_phonon,chow_girvin_exp}, since the
typical values of wave vectors and frequencies of the phonons are
small compared to the scales over which the inverse conductivity
changes, we replace $G \propto \omega$. Again a simple  power
counting results in $P \sim T^4$ akin to what is seen in the IQHE
phases.

A real calculation of $P$ requires a knowledge of the finite
temperature conductivity. At low enough temperatures, topological
defects are thermally excited and they constitute the primary
contribution to the ultra-low frequency conductivity. Though  our
formalism cannot treat the defects, we present heuristic arguments
to justify our above result. A reasonable hypothesis for the low
temperature frequency dependent conductivity, is
$\sigma(q,\omega,T) \sim \sigma(q,\omega,0) +
\sigma_d(q,\omega,T)$, where $\sigma_d$ is the defect
contribution. The presence of $\sigma_d$ leads to a finite but
small dc conductivity $\sigma(T)$ in the long wavelength limit. We
expect this to be the dominant contribution for small frequencies
$\omega < \omega_d$  and for frequencies  $\omega > \omega_d$, we
expect the finite frequency contribution to dominate. The cut-off
frequency $\omega_d$ is fixed by $\sigma(\omega,T=0)\sim
\sigma_d(T)$. Since for small frequencies, $\sigma(\omega,T=0)\sim
A \omega^2$, we obtain $\omega_d = \sqrt{\sigma_d(T)/A}$. This
leads to the form that  ${\rm Re}(1/\sigma( q,\omega,T) \sim
\frac1{\sigma_d(T)} + Re (1/\sigma(q,\omega,T=0)$. Using this form
in (\ref{power}), and doing the usual power counting, we obtain $P
\propto T^4 + a T \sqrt{\sigma_d(T)}$. Since in the limit of low
temperatures $\sigma_d(T)$ is expected to be exponentially small,
the power spectrum is dominated by the $T^4$ term, thereby
justifying our naive analysis of the preceding paragraph. This
coincidence is related to the fact that both the pinned crystal
and the IQHE phases are incompressible. Clearly, the variations
seen in the inverse conductivity will modify this result, but we
expect the result to hold for some range of low enough
temperatures with a crossover to a different scaling form at
higher temperatures.

\section {Conclusions and experimental propositions} \label{sec:data}

Before we conclude, we delve into the possibility of obtaining
quantitative estimates for the theoretical scales $R_a$ and
$\Sigma$  in the so called WC phase. These are very important for
a real comparison of theoretical predictions with experimental
data. The available experimental data which includes mobility and
resistivity measurements, permits one to calculate basic
quantities like the the fermi momentum $k_F$ and  scattering
rates. The fermi momentum can be  expressed as a simple function
of density $\pi a^2=n$ or $r_s=a/a_B$, the mass renormalization
factor $m_r$ and the relative permittivity $\epsilon_r$ of the
material. The Bohr radius of the electron in the given material is
$a_B= \epsilon_r/m_r  0.53 A^o$. Using the relation for the Fermi
energy of the electron gas, $E_F= {{\hbar}^2\over {m^* a^2}}\equiv
{\hbar}^2 k_F^2/m^*$, one finds
\begin{equation}
 k_F=1.768 n^{\frac12} \simeq {{m_r}\over {\epsilon_r r_s}} 1.88
 \times 10^{10}
\end{equation}
$k_F$ is expressed in units of inverse meters.
We also find that the plasma frequency of the electron gas
\begin{equation}
\omega_{plas}^2= \frac{e^2}{2\pi m^* \epsilon a^3}
\label{plasma}
\end{equation}
and the pinning frequency $\omega_p= \sqrt{\frac{c_T}{\rho_m}}
R_c^{-1}$  which as shown earlier, sets the scale for the
conductivity and other dynamical quantities, satisfy the following
relation
\begin{equation}
 \frac{\omega_{plas}}{\omega_p}\sim   \frac{R_c}{\sqrt{2}a}
\end{equation}
where, the Larkin length $R_c \simeq  R_a [\xi_0/a]^3$. Since most
experiments on the 2D electron gas in Si MOSFETS and clean
heterojunctions  probe dc quantities, they cannot permit a simple
extraction of the disorder scales.  To use the available data on
dc transport to extract the disorder scales, would require a
detailed theoretical study of finite temperature dc transport,
which is clearly beyond the scope of this paper and that of the
variational method.  The only other possibility is through
measurements of microwave   conductivity  which presents a
straightforward way of extracting the pinning frequency.  These
measurements in conjunction with SAW measurements which can be
used to obtain the effective elastic modulus  can then be combined
with theoretical studies  to verify the collectively pinned nature
of the crystal.

To conclude, in this paper, we have studied the the physical
properties of the Wigner crystal of a low density electron gas
that is pinned by weak gaussian disorder. We have used the
standard combination of the gaussian variational method and
replicas to calculate a host of transport properties. We presented
results for the evolution of the dynamical conductivity as a
function of density in both the regimes where the effective
particle size $\xi$ is bigger or smaller than the disorder
correlation length $r_f$, the  SAW
attenuation and the putative power radiated by the pinned crystal
to its environment. We hope that our results provide a stimulus
for microwave conductivity measurements which have been
instrumental in establishing the crystalline nature of the 2DEG in
very strong magnetic
fields\cite{li_conductivity_wigner_magneticfield}.  This will then
instigate a real debate as to whether the insulating phase seen in
systems which exhibit the 2d MIT is really a Wigner crystal or
some other phase.

Clearly, many a question needs to be addressed
in order to understand the real nature of the  crystalline state.
As mentioned earlier, one first needs to evaluate the elastic
constants of the crystal as a function of the density. The
variational wave function used in the \sref{sec:varxi} to obtain
the effective particle size can be used to obtain reasonable
estimates of these constants. We expect the bulk modulus to not
differ much from the result of
Ref.~\onlinecite{bonsall_elastic_wigner} since this is fixed
solely by the coulomb interaction. However, the shear modulus is
bound to deviate from the classical value especially as it
approaches melting when the density is varied. These values when
plugged into the results obtained in this paper can lead to a
behavior, not too far from the melting of the crystal, which
deviates strongly from that expected for a system with the
classical elastic constants. Secondly, in the absence of a strong
magnetic field, the spinful nature of the electrons might lead to
interesting spin physics in the crystalline phase. Interesting
spin liquid behaviors have been seen in studies of the multi-
exchange spin models in a triangular lattice
\cite{misguich_ringexchange} and it would be pertinent to ask if
such exotic spin physics arises in the crystal and how does this
change when the system is pinned by the incipient disorder.
Moreover, any theoretical study of the magneto-resistance needs to
take the spin physics into account. Other questions concern the
behavior of the system close to depinning as also the dc transport
at finite temperature.  Some of these questions will be addressed
in future work.

\begin{acknowledgments}
This work was supported in part be the Swiss National Fund through
Manep and Division II.
\end{acknowledgments}

\appendix

\section{RPA} \label{ap:rpacoul}

A standard way to compute the compressibility of a charged system
is through the free energy $F$ of the system  assuming neutrality,
and from that to compute $d \mu/d N$. Since the system is always
supposed to remain neutral the compressibility stays finite even
for a charged system. However, this method is cumbersome since the
calculation of the free energy is usually more difficult than that
of a correlation function. In this appendix we show, using the RPA
approximation, that the formula (\ref{eq:congen}) indeed yields to
the standard results for the compressibility of a normal metal.

Using (\ref{eq:hamdep}) it is easy to check that the
susceptibilities $\chi_{\alpha\beta} = \langle \rho_\alpha
\rho_\beta \rangle$ are given, in RPA, by
\begin{equation} \label{eq:rpaeq}
 \left(\begin{array}{c} \chi_{11} \\ \chi_{12} \end{array} \right)
 =
 - \left(\begin{array}{cc} \chi^0 V_{11} & \chi^0 V_{12} \\
                         \chi^0 V_{12} & \chi^0 V_{11}
       \end{array} \right)
 \left(\begin{array}{c} \chi_{11} \\ \chi_{12} \end{array} \right)
 +
 \left(\begin{array}{c} \chi^0 \\ 0 \end{array} \right)
\end{equation}
where $\chi^0$ is the bare (i.e. for $H^0$ only) density-density
correlation function in one of the systems and  the $q$
and $\omega$ dependence is implicit. It is easy to solve
(\ref{eq:rpaeq}) to obtain the (q dependent) capacitance
\begin{equation}
 \chi_{11}(\omega,{\bf q}) - \chi_{12}(\omega,{\bf q}) =
 \frac{\chi^0(\omega,{\bf q})}{1 + \chi^0(\omega,{\bf q})(V_{11}({\bf q}) -
 V_{12}({\bf q}))}
\end{equation}
The Fourier transform of the Coulomb potentials are given by
(\ref{eq:cancelcoul}). The true capacitance corresponds to $\omega
= 0$ and the limit $q\to 0$ which leads to
\begin{equation}
 C = \frac1{(2\chi^0(q=0))^{-1} + 4 \pi d}
\end{equation}
The total capacitance is thus the sum of a geometrical one $C_{\rm
geom}$ and one due to the electron gas residing in the planes
$C_{\rm el}$
\begin{equation} \label{eq:geom}
 \frac1{C} = \frac1{C_{\rm geom}} + \frac1{C_{\rm el}}
\end{equation}
The geometrical one is the standard $1/(4\pi d)$ result. For a
simple electron gas, $\chi^0(q=0)^{-1}$ is simply the screening
length $\lambda$. One thus recovers that the
geometrical distance $d$ between the planes is effectively  enhanced  by the
electronic screening length $\lambda$ on each plane.

\section{Calculation of the elastic coefficients} \label{ap:bonsder}

The decomposition of the density (\ref{eq:fourdens}) allows to
recover quite simply the formulas of
Ref.~\onlinecite{bonsall_elastic_wigner} for the elastic
coefficients. Indeed if one assumes that the interaction between
the particles of the crystal is
\begin{equation} \label{eq:intbas}
 H = \frac12 \int_{{\bf r},{\bf r}'} V({\bf r}-{\bf r}')[\rho({\bf r}) -
 \rho_0][\rho({\bf r}')-\rho_0]
\end{equation}
where $V({\bf r}-{\bf r}')$ is the (three dimensional) Coulomb interaction.
The decomposition of the density (\ref{eq:fourdens}) allows us
to rewrite  (\ref{eq:intbas})  as
\begin{multline} \label{eq:interdens}
 H = \frac{\rho_0^2}2 \int_{{\bf r},{\bf r}'} V({\bf r}-{\bf r}') [\nabla\cdot {\bf u}({\bf r})][\nabla\cdot
 {\bf u}({\bf r}')] + \\
 \frac{\rho_0^2}2 \sum_{{\bf K}\ne 0} \int_{{\bf r},{\bf r}'} V({\bf r}-{\bf r}') e^{i{\bf K}\cdot ({\bf r}-{\bf r}')}e^{-i{\bf K}\cdot
 ({\bf u}({\bf r})-{\bf u}({\bf r}'))}
\end{multline}
where we have only kept terms that are not averaged to zero due to
the translational invariance. The first term in
(\ref{eq:interdens}) gives back the term proportional to $q$ in
(\ref{moduli}). Indeed one obtains
\begin{equation}
 H_1 = \frac1{2\Omega} \sum_{\bf q} \frac{e^2 \rho_0^2 q}{\epsilon}
 u_L^*({\bf q}) u_L({\bf q})
\end{equation}
This gives the value $d = \rho_c^2/\epsilon$.

The second term can be expanded to second order in $u$ to give the
elastic coefficients. The first order term vanishes because the
perfect lattice ($u=0$) is an extremum of the energy. Up to second
order in $u$,
\begin{multline} \label{eq:h2}
 H_2 = \frac{\rho_0^2}2 \sum_{{\bf K}\ne 0} \int_{{\bf r},{\bf r}'} V({\bf r}-{\bf r}') e^{i{\bf K}\cdot ({\bf r}-{\bf r}')}
 K_\alpha K_\beta \\
 (u_\alpha({\bf r})-u_\alpha({\bf r}'))(u_\beta({\bf r})-u_\beta({\bf r}'))
\end{multline}
where the summation on the coordinate indices $\alpha,\beta =
(x,y)$ is implicit. Using
\begin{equation}
 u_\alpha({\bf r})-u_\alpha({\bf r}') = \frac1\Omega \sum_{\bf q} u_\alpha({\bf q})(e^{i {\bf q} \cdot
{\bf  r}} - e^{i {\bf q }\cdot {\bf r}'})
\end{equation}
One can rewrite (\ref{eq:h2}) as
\begin{multline}
 H_2 = \frac{\rho_0^2}{2\Omega^2} \sum_{{\bf K}\ne 0} \sum_{{\bf q},{\bf q}'}\int_{{\bf r},{\bf r}'} V({\bf r}-{\bf r}') e^{i{\bf K}\cdot ({\bf r}-{\bf r}')}
 K_\alpha K_\beta \\ u_\alpha({\bf q}) u_\beta({\bf q}')
 (e^{i {\bf q}\cdot  {\bf r}} - e^{i {\bf q}\cdot  {\bf r}'}) (e^{i {\bf q}'\cdot  {\bf r}} - e^{i {\bf q}'\cdot {\bf r}'})\\
\end{multline}
Using center of mass ${\bf R}={\bf r}+{\bf r}'$ and relative ${\bf
r}_0 = {\bf r}-{\bf r}'$ coordinates can be rewritten as
\begin{multline} \label{eq:h2new}
 H_2 = \frac{\rho_0^2}{2\Omega} \sum_{{\bf K}\ne 0} \sum_{{\bf q}}\int_{{\bf r}_0} V({\bf r}_0) e^{i{\bf K}\cdot  {\bf r}_0}
 K_\alpha K_\beta u_\alpha({\bf q}) u_\beta(-{\bf q}) \\
 (2 -2 \cos({\bf q}\cdot  {\bf r}_0))\\
\end{multline}
which using the Fourier transform of the potential $V(r_0)$
denoted $\tilde V({\bf q}) = 1/q$ one obtains
\begin{widetext}
\begin{equation}
 H_2 = \frac{\rho_0^2}{2\Omega} \sum_{{\bf q}}
 [\sum_{{\bf K}\ne 0} K_\alpha K_\beta (2 \tilde V({\bf K}) - \tilde V({\bf K}+{\bf q}) - \tilde V({\bf K}-{\bf q})] u_\alpha({\bf q}) u_\beta(-{\bf q})
\end{equation}
\end{widetext}
which gives the classical elastic constants.

\begin{thebibliography}{43}
\expandafter\ifx\csname natexlab\endcsname\relax\def\natexlab#1{#1}\fi
\expandafter\ifx\csname bibnamefont\endcsname\relax
  \def\bibnamefont#1{#1}\fi
\expandafter\ifx\csname bibfnamefont\endcsname\relax
  \def\bibfnamefont#1{#1}\fi
\expandafter\ifx\csname citenamefont\endcsname\relax
  \def\citenamefont#1{#1}\fi
\expandafter\ifx\csname url\endcsname\relax
  \def\url#1{\texttt{#1}}\fi
\expandafter\ifx\csname urlprefix\endcsname\relax\def\urlprefix{URL }\fi
\providecommand{\bibinfo}[2]{#2}
\providecommand{\eprint}[2][]{\url{#2}}

\bibitem[{\citenamefont{Abrahams et~al.}(2001)\citenamefont{Abrahams,
  Kravchenko, and Sarachik}}]{abrahams_review_mit_2d}
\bibinfo{author}{\bibfnamefont{E.}~\bibnamefont{Abrahams}},
  \bibinfo{author}{\bibfnamefont{S.~V.} \bibnamefont{Kravchenko}},
  \bibnamefont{and} \bibinfo{author}{\bibfnamefont{M.~P.}
  \bibnamefont{Sarachik}}, \bibinfo{journal}{Rev. Mod. Phys.}
  \textbf{\bibinfo{volume}{73}}, \bibinfo{pages}{251} (\bibinfo{year}{2001}).

\bibitem[{\citenamefont{Anderson}(1958)}]{anderson_localisation}
\bibinfo{author}{\bibfnamefont{P.~W.} \bibnamefont{Anderson}},
  \bibinfo{journal}{Phys. Rev.} \textbf{\bibinfo{volume}{109}},
  \bibinfo{pages}{1492} (\bibinfo{year}{1958}).

\bibitem[{\citenamefont{Abrahams et~al.}(1979)\citenamefont{Abrahams, Anderson,
  Licciardello, and Ramakrishnan}}]{abrahams_loc}
\bibinfo{author}{\bibfnamefont{E.}~\bibnamefont{Abrahams}},
  \bibinfo{author}{\bibfnamefont{P.~W.} \bibnamefont{Anderson}},
  \bibinfo{author}{\bibfnamefont{D.~C.} \bibnamefont{Licciardello}},
  \bibnamefont{and} \bibinfo{author}{\bibfnamefont{T.~V.}
  \bibnamefont{Ramakrishnan}}, \bibinfo{journal}{Phys. Rev. Lett.}
  \textbf{\bibinfo{volume}{42}}, \bibinfo{pages}{673} (\bibinfo{year}{1979}).

\bibitem[{\citenamefont{Lee and Ramakhrishnan}(1985)}]{lee_mit_long}
\bibinfo{author}{\bibfnamefont{P.~A.} \bibnamefont{Lee}} \bibnamefont{and}
  \bibinfo{author}{\bibfnamefont{T.~V.} \bibnamefont{Ramakhrishnan}},
  \bibinfo{journal}{Rev. Mod. Phys.} \textbf{\bibinfo{volume}{57}},
  \bibinfo{pages}{287} (\bibinfo{year}{1985}).

\bibitem[{\citenamefont{Finkelstein}(1984)}]{finkelstein_localization_interact%
ions}
\bibinfo{author}{\bibfnamefont{A.~M.} \bibnamefont{Finkelstein}},
  \bibinfo{journal}{Z. Phys. B} \textbf{\bibinfo{volume}{56}},
  \bibinfo{pages}{189} (\bibinfo{year}{1984}).

\bibitem[{bel()}]{belitz_localization_review}
\bibinfo{note}{For a review see: D. Belitz and T. R. Kirkpatrick, Rev. Mod.
  Phys. {\bf 66} 261 (1994).}

\bibitem[{\citenamefont{Gogolin et~al.}(1999)\citenamefont{Gogolin, Nersesyan,
  and Tsvelik}}]{gogolin_1dbook}
\bibinfo{author}{\bibfnamefont{A.~O.} \bibnamefont{Gogolin}},
  \bibinfo{author}{\bibfnamefont{A.~A.} \bibnamefont{Nersesyan}},
  \bibnamefont{and} \bibinfo{author}{\bibfnamefont{A.~M.}
  \bibnamefont{Tsvelik}}, \emph{\bibinfo{title}{Bosonization and Strongly
  Correlated Systems}} (\bibinfo{publisher}{Cambridge University Press},
  \bibinfo{address}{Cambridge}, \bibinfo{year}{1999}).

\bibitem[{\citenamefont{Giamarchi}(2004)}]{giamarchi_book_1d}
\bibinfo{author}{\bibfnamefont{T.}~\bibnamefont{Giamarchi}},
  \emph{\bibinfo{title}{Quantum Physics in One Dimension}}
  (\bibinfo{publisher}{Oxford University Press}, \bibinfo{address}{Oxford},
  \bibinfo{year}{2004}).

\bibitem[{\citenamefont{Giamarchi and Schulz}(1988)}]{giamarchi_loc}
\bibinfo{author}{\bibfnamefont{T.}~\bibnamefont{Giamarchi}} \bibnamefont{and}
  \bibinfo{author}{\bibfnamefont{H.~J.} \bibnamefont{Schulz}},
  \bibinfo{journal}{Phys. Rev. B} \textbf{\bibinfo{volume}{37}},
  \bibinfo{pages}{325} (\bibinfo{year}{1988}).

\bibitem[{\citenamefont{Wigner}(1934)}]{wigner_crystal}
\bibinfo{author}{\bibfnamefont{E.}~\bibnamefont{Wigner}},
  \bibinfo{journal}{Phys. Rev.} \textbf{\bibinfo{volume}{46}},
  \bibinfo{pages}{1002} (\bibinfo{year}{1934}).

\bibitem[{\citenamefont{Ceperley}(1984)}]{ceperley_qmc_wigner}
\bibinfo{author}{\bibfnamefont{D.}~\bibnamefont{Ceperley}},
  \bibinfo{journal}{Phys. Rev. B} \textbf{\bibinfo{volume}{46}},
  \bibinfo{pages}{1002} (\bibinfo{year}{1984}).

\bibitem[{\citenamefont{Fukuyama and
  Lee}(1978{\natexlab{a}})}]{fukuyama_cdw_magnetic}
\bibinfo{author}{\bibfnamefont{H.}~\bibnamefont{Fukuyama}} \bibnamefont{and}
  \bibinfo{author}{\bibfnamefont{P.}~\bibnamefont{Lee}},
  \bibinfo{journal}{Phys. Rev. B} \textbf{\bibinfo{volume}{18}},
  \bibinfo{pages}{6245} (\bibinfo{year}{1978}{\natexlab{a}}).

\bibitem[{\citenamefont{Blatter et~al.}(1994)\citenamefont{Blatter, Feigel'man,
  Geshkenbein, Larkin, and Vinokur}}]{blatter_vortex_review}
\bibinfo{author}{\bibfnamefont{G.}~\bibnamefont{Blatter}},
  \bibinfo{author}{\bibfnamefont{M.~V.} \bibnamefont{Feigel'man}},
  \bibinfo{author}{\bibfnamefont{V.~B.} \bibnamefont{Geshkenbein}},
  \bibinfo{author}{\bibfnamefont{A.~I.} \bibnamefont{Larkin}},
  \bibnamefont{and} \bibinfo{author}{\bibfnamefont{V.~M.}
  \bibnamefont{Vinokur}}, \bibinfo{journal}{Rev. Mod. Phys.}
  \textbf{\bibinfo{volume}{66}}, \bibinfo{pages}{1125} (\bibinfo{year}{1994}).

\bibitem[{\citenamefont{Nattermann and
  Scheidl}(2000)}]{nattermann_vortex_review}
\bibinfo{author}{\bibfnamefont{T.}~\bibnamefont{Nattermann}} \bibnamefont{and}
  \bibinfo{author}{\bibfnamefont{S.}~\bibnamefont{Scheidl}},
  \bibinfo{journal}{Adv. Phys.} \textbf{\bibinfo{volume}{49}},
  \bibinfo{pages}{607} (\bibinfo{year}{2000}).

\bibitem[{\citenamefont{Giamarchi and
  Bhattacharya}(2002)}]{giamarchi_vortex_review}
\bibinfo{author}{\bibfnamefont{T.}~\bibnamefont{Giamarchi}} \bibnamefont{and}
  \bibinfo{author}{\bibfnamefont{S.}~\bibnamefont{Bhattacharya}}, in
  \emph{\bibinfo{booktitle}{High Magnetic Fields: Applications in Condensed
  Matter Physics and Spectroscopy}}, edited by
  \bibinfo{editor}{\bibfnamefont{C.}~\bibnamefont{{Berthier {\it et al.}}}}
  (\bibinfo{publisher}{Springer-Verlag}, \bibinfo{address}{Berlin},
  \bibinfo{year}{2002}), p. \bibinfo{pages}{314},
  \bibinfo{note}{cond-mat/0111052}.

\bibitem[{\citenamefont{Giamarchi and
  Orignac}(2003)}]{giamarchi_quantum_pinning}
\bibinfo{author}{\bibfnamefont{T.}~\bibnamefont{Giamarchi}} \bibnamefont{and}
  \bibinfo{author}{\bibfnamefont{E.}~\bibnamefont{Orignac}}, in
  \emph{\bibinfo{booktitle}{Theoretical Methods for Strongly Correlated
  Electrons}}, edited by
  \bibinfo{editor}{\bibfnamefont{D.}~\bibnamefont{{S{\'e}nechal {\it et al.}}}}
  (\bibinfo{publisher}{Springer}, \bibinfo{address}{New York},
  \bibinfo{year}{2003}), CRM Series in Mathematical Physics,
  \bibinfo{note}{cond-mat/0005220}.

\bibitem[{\citenamefont{Giamarchi}(2003)}]{giamarchi_varenna_wigner_review}
\bibinfo{author}{\bibfnamefont{T.}~\bibnamefont{Giamarchi}},
  \emph{\bibinfo{title}{Quantum phenomena in mesoscopic system}}
  (\bibinfo{publisher}{IOS Press}, \bibinfo{address}{Amsterdam},
  \bibinfo{year}{2003}), \bibinfo{note}{cond-mat/0403531}.

\bibitem[{\citenamefont{Normand et~al.}(1992)\citenamefont{Normand, Littlewood,
  and Millis}}]{normand_millis_wigner}
\bibinfo{author}{\bibfnamefont{B.~G.~A.} \bibnamefont{Normand}},
  \bibinfo{author}{\bibfnamefont{P.~B.} \bibnamefont{Littlewood}},
  \bibnamefont{and} \bibinfo{author}{\bibfnamefont{A.~J.}
  \bibnamefont{Millis}}, \bibinfo{journal}{Phys. Rev. B}
  \textbf{\bibinfo{volume}{46}}, \bibinfo{pages}{3920} (\bibinfo{year}{1992}).

\bibitem[{\citenamefont{Chitra et~al.}(1998)\citenamefont{Chitra, Giamarchi,
  and {Le Doussal}}}]{chitra_wigner_hall}
\bibinfo{author}{\bibfnamefont{R.}~\bibnamefont{Chitra}},
  \bibinfo{author}{\bibfnamefont{T.}~\bibnamefont{Giamarchi}},
  \bibnamefont{and} \bibinfo{author}{\bibfnamefont{P.}~\bibnamefont{{Le
  Doussal}}}, \bibinfo{journal}{Phys. Rev. Lett.}
  \textbf{\bibinfo{volume}{80}}, \bibinfo{pages}{3827} (\bibinfo{year}{1998}).

\bibitem[{\citenamefont{Yi and Fertig}(2000)}]{yi_pinning_wigner}
\bibinfo{author}{\bibfnamefont{H.~M.} \bibnamefont{Yi}} \bibnamefont{and}
  \bibinfo{author}{\bibfnamefont{H.~A.} \bibnamefont{Fertig}},
  \bibinfo{journal}{Phys. Rev. B} \textbf{\bibinfo{volume}{61}},
  \bibinfo{pages}{5311} (\bibinfo{year}{2000}).

\bibitem[{\citenamefont{Chitra et~al.}(2001)\citenamefont{Chitra, Giamarchi,
  and {Le Doussal}}}]{chitra_wigner_long}
\bibinfo{author}{\bibfnamefont{R.}~\bibnamefont{Chitra}},
  \bibinfo{author}{\bibfnamefont{T.}~\bibnamefont{Giamarchi}},
  \bibnamefont{and} \bibinfo{author}{\bibfnamefont{P.}~\bibnamefont{{Le
  Doussal}}}, \bibinfo{journal}{Phys. Rev. B} \textbf{\bibinfo{volume}{65}},
  \bibinfo{pages}{035312} (\bibinfo{year}{2001}).

\bibitem[{\citenamefont{Fogler and Huse}(2000)}]{fogler_pinning_wigner}
\bibinfo{author}{\bibfnamefont{M.~M.} \bibnamefont{Fogler}} \bibnamefont{and}
  \bibinfo{author}{\bibfnamefont{D.~A.} \bibnamefont{Huse}},
  \bibinfo{journal}{Phys. Rev. B} \textbf{\bibinfo{volume}{62}},
  \bibinfo{pages}{7553} (\bibinfo{year}{2000}).

\bibitem[{\citenamefont{Li et~al.}(1997)\citenamefont{Li, Engel, Shahar, Tsui,
  and Shayegan}}]{li_conductivity_wigner_magneticfield}
\bibinfo{author}{\bibfnamefont{C.~C.} \bibnamefont{Li}},
  \bibinfo{author}{\bibfnamefont{L.~W.} \bibnamefont{Engel}},
  \bibinfo{author}{\bibfnamefont{D.}~\bibnamefont{Shahar}},
  \bibinfo{author}{\bibfnamefont{D.~C.} \bibnamefont{Tsui}}, \bibnamefont{and}
  \bibinfo{author}{\bibfnamefont{M.}~\bibnamefont{Shayegan}},
  \bibinfo{journal}{Phys. Rev. Lett.} \textbf{\bibinfo{volume}{79}},
  \bibinfo{pages}{1353} (\bibinfo{year}{1997}).

\bibitem[{\citenamefont{Li et~al.}(2000)\citenamefont{Li, Yoon, Engel, Shahar,
  Tsui, and Shayegan}}]{li_conductivity_wigner_density}
\bibinfo{author}{\bibfnamefont{C.~C.} \bibnamefont{Li}},
  \bibinfo{author}{\bibfnamefont{J.}~\bibnamefont{Yoon}},
  \bibinfo{author}{\bibfnamefont{L.~W.} \bibnamefont{Engel}},
  \bibinfo{author}{\bibfnamefont{D.}~\bibnamefont{Shahar}},
  \bibinfo{author}{\bibfnamefont{D.~C.} \bibnamefont{Tsui}}, \bibnamefont{and}
  \bibinfo{author}{\bibfnamefont{M.}~\bibnamefont{Shayegan}},
  \bibinfo{journal}{Phys. Rev. B} \textbf{\bibinfo{volume}{61}},
  \bibinfo{pages}{10905} (\bibinfo{year}{2000}).

\bibitem[{\citenamefont{Bonsall and Maradudin}(1977)}]{bonsall_elastic_wigner}
\bibinfo{author}{\bibfnamefont{L.}~\bibnamefont{Bonsall}} \bibnamefont{and}
  \bibinfo{author}{\bibfnamefont{A.~A.} \bibnamefont{Maradudin}},
  \bibinfo{journal}{Phys. Rev. B} \textbf{\bibinfo{volume}{15}},
  \bibinfo{pages}{1959} (\bibinfo{year}{1977}).

\bibitem[{\citenamefont{Giamarchi and {Le
  Doussal}}(1994)}]{giamarchi_vortex_short}
\bibinfo{author}{\bibfnamefont{T.}~\bibnamefont{Giamarchi}} \bibnamefont{and}
  \bibinfo{author}{\bibfnamefont{P.}~\bibnamefont{{Le Doussal}}},
  \bibinfo{journal}{Phys. Rev. Lett.} \textbf{\bibinfo{volume}{72}},
  \bibinfo{pages}{1530} (\bibinfo{year}{1994}).

\bibitem[{\citenamefont{Giamarchi and {Le
  Doussal}}(1995)}]{giamarchi_vortex_long}
\bibinfo{author}{\bibfnamefont{T.}~\bibnamefont{Giamarchi}} \bibnamefont{and}
  \bibinfo{author}{\bibfnamefont{P.}~\bibnamefont{{Le Doussal}}},
  \bibinfo{journal}{Phys. Rev. B} \textbf{\bibinfo{volume}{52}},
  \bibinfo{pages}{1242} (\bibinfo{year}{1995}).

\bibitem[{\citenamefont{Giamarchi and {Le
  Doussal}}(1996)}]{giamarchi_columnar_variat}
\bibinfo{author}{\bibfnamefont{T.}~\bibnamefont{Giamarchi}} \bibnamefont{and}
  \bibinfo{author}{\bibfnamefont{P.}~\bibnamefont{{Le Doussal}}},
  \bibinfo{journal}{Phys. Rev. B} \textbf{\bibinfo{volume}{53}},
  \bibinfo{pages}{15206} (\bibinfo{year}{1996}).

\bibitem[{\citenamefont{Giamarchi and {Le
  Doussal}}(1998)}]{giamarchi_book_young}
\bibinfo{author}{\bibfnamefont{T.}~\bibnamefont{Giamarchi}} \bibnamefont{and}
  \bibinfo{author}{\bibfnamefont{P.}~\bibnamefont{{Le Doussal}}},
  \emph{\bibinfo{title}{Statics and dynamics of disordered elastic systems}}
  (\bibinfo{publisher}{World Scientific}, \bibinfo{address}{Singapore},
  \bibinfo{year}{1998}), p. \bibinfo{pages}{321},
  \bibinfo{note}{cond-mat/9705096}.

\bibitem[{\citenamefont{Eisenstein et~al.}(1992)\citenamefont{Eisenstein,
  Pfeiffer, and West}}]{eisenstein_hall_compressibility}
\bibinfo{author}{\bibfnamefont{J.~P.} \bibnamefont{Eisenstein}},
  \bibinfo{author}{\bibfnamefont{L.~N.} \bibnamefont{Pfeiffer}},
  \bibnamefont{and} \bibinfo{author}{\bibfnamefont{K.~W.} \bibnamefont{West}},
  \bibinfo{journal}{Phys. Rev. Lett.} \textbf{\bibinfo{volume}{68}},
  \bibinfo{pages}{674} (\bibinfo{year}{1992}).

\bibitem[{\citenamefont{Giamarchi}(2002)}]{giamarchi_wigner_review}
\bibinfo{author}{\bibfnamefont{T.}~\bibnamefont{Giamarchi}}, in
  \emph{\bibinfo{booktitle}{Strongly correlated fermions and bosons in low
  dimensional disordered systems}}, edited by
  \bibinfo{editor}{\bibfnamefont{I.~V.} \bibnamefont{{Lerner {\it et al.}}}}
  (\bibinfo{publisher}{Kluwer}, \bibinfo{address}{Dordrecht},
  \bibinfo{year}{2002}), \bibinfo{note}{cond-mat/0205099}.

\bibitem[{\citenamefont{Falko}(1994)}]{falko_bilayer}
\bibinfo{author}{\bibfnamefont{V.~I.} \bibnamefont{Falko}},
  \bibinfo{journal}{Phys. Rev. B} \textbf{\bibinfo{volume}{49}},
  \bibinfo{pages}{7774} (\bibinfo{year}{1994}).

\bibitem[{\citenamefont{Nguyen et~al.}(2001)\citenamefont{Nguyen, Grosberg, and
  Shklovskii}}]{nguyen_wigner_biology_houches}
\bibinfo{author}{\bibfnamefont{T.~T.} \bibnamefont{Nguyen}},
  \bibinfo{author}{\bibfnamefont{A.~Y.} \bibnamefont{Grosberg}},
  \bibnamefont{and} \bibinfo{author}{\bibfnamefont{B.~I.}
  \bibnamefont{Shklovskii}} (\bibinfo{year}{2001}),
  \bibinfo{note}{cond-mat/0101103}.

\bibitem[{\citenamefont{{Li {\it et al.}}}(1995)}]{li_optical_wigner}
\bibinfo{author}{\bibfnamefont{Y.~P.} \bibnamefont{{Li {\it et al.}}}},
  \bibinfo{journal}{Solid State Commun.} \textbf{\bibinfo{volume}{95}},
  \bibinfo{pages}{619} (\bibinfo{year}{1995}).

\bibitem[{\citenamefont{Andrei et~al.}(1988)\citenamefont{Andrei, Deville,
  Glattli, and Williams}}]{andrei_saw}
\bibinfo{author}{\bibfnamefont{E.~Y.} \bibnamefont{Andrei}},
  \bibinfo{author}{\bibfnamefont{G.}~\bibnamefont{Deville}},
  \bibinfo{author}{\bibfnamefont{D.~C.} \bibnamefont{Glattli}},
  \bibnamefont{and} \bibinfo{author}{\bibfnamefont{F.~I.~B.}
  \bibnamefont{Williams}}, \bibinfo{journal}{Phys. Rev. Lett.}
  \textbf{\bibinfo{volume}{60}}, \bibinfo{pages}{2765} (\bibinfo{year}{1988}).

\bibitem[{\citenamefont{Williams et~al.}(1991)\citenamefont{Williams, Wright,
  Clark, Andrei, Deville, Glattli, Probst, Etienne, Dorin, Foxon
  et~al.}}]{williams_wigner_threshold}
\bibinfo{author}{\bibfnamefont{F.~I.~B.} \bibnamefont{Williams}},
  \bibinfo{author}{\bibfnamefont{P.~A.} \bibnamefont{Wright}},
  \bibinfo{author}{\bibfnamefont{R.~G.} \bibnamefont{Clark}},
  \bibinfo{author}{\bibfnamefont{E.~Y.} \bibnamefont{Andrei}},
  \bibinfo{author}{\bibfnamefont{G.}~\bibnamefont{Deville}},
  \bibinfo{author}{\bibfnamefont{D.~C.} \bibnamefont{Glattli}},
  \bibinfo{author}{\bibfnamefont{O.}~\bibnamefont{Probst}},
  \bibinfo{author}{\bibfnamefont{B.}~\bibnamefont{Etienne}},
  \bibinfo{author}{\bibfnamefont{C.}~\bibnamefont{Dorin}},
  \bibinfo{author}{\bibfnamefont{C.~T.} \bibnamefont{Foxon}},
  \bibnamefont{et~al.}, \bibinfo{journal}{Phys. Rev. Lett.}
  \textbf{\bibinfo{volume}{66}}, \bibinfo{pages}{3285} (\bibinfo{year}{1991}).

\bibitem[{\citenamefont{Fukuyama and
  Lee}(1978{\natexlab{b}})}]{fukuyama_pinning}
\bibinfo{author}{\bibfnamefont{H.}~\bibnamefont{Fukuyama}} \bibnamefont{and}
  \bibinfo{author}{\bibfnamefont{P.~A.} \bibnamefont{Lee}},
  \bibinfo{journal}{Phys. Rev. B} \textbf{\bibinfo{volume}{17}},
  \bibinfo{pages}{535} (\bibinfo{year}{1978}{\natexlab{b}}).

\bibitem[{\citenamefont{Larkin and
  Ovchinnikov}(1979)}]{larkin_ovchinnikov_pinning}
\bibinfo{author}{\bibfnamefont{A.~I.} \bibnamefont{Larkin}} \bibnamefont{and}
  \bibinfo{author}{\bibfnamefont{Y.~N.} \bibnamefont{Ovchinnikov}},
  \bibinfo{journal}{J. Low Temp. Phys} \textbf{\bibinfo{volume}{34}},
  \bibinfo{pages}{409} (\bibinfo{year}{1979}).

\bibitem[{\citenamefont{Halperin et~al.}(1993)\citenamefont{Halperin, Lee, and
  Read}}]{halperin_lee_saw}
\bibinfo{author}{\bibfnamefont{B.~I.} \bibnamefont{Halperin}},
  \bibinfo{author}{\bibfnamefont{P.~A.} \bibnamefont{Lee}}, \bibnamefont{and}
  \bibinfo{author}{\bibfnamefont{N.}~\bibnamefont{Read}},
  \bibinfo{journal}{Phys. Rev. B} \textbf{\bibinfo{volume}{47}},
  \bibinfo{pages}{7312} (\bibinfo{year}{1993}).

\bibitem[{\citenamefont{Chow et~al.}(1996)\citenamefont{Chow, Wei, Girvin, and
  Shayegan}}]{chow_girvin_phonon}
\bibinfo{author}{\bibfnamefont{E.}~\bibnamefont{Chow}},
  \bibinfo{author}{\bibfnamefont{H.~P.} \bibnamefont{Wei}},
  \bibinfo{author}{\bibfnamefont{S.~M.} \bibnamefont{Girvin}},
  \bibnamefont{and} \bibinfo{author}{\bibfnamefont{M.}~\bibnamefont{Shayegan}},
  \bibinfo{journal}{Phys. Rev. Lett.} \textbf{\bibinfo{volume}{77}},
  \bibinfo{pages}{1143} (\bibinfo{year}{1996}).

\bibitem[{\citenamefont{Chow et~al.}(1997)\citenamefont{Chow, Wei, Girvin, Jan,
  and Cunningham}}]{chow_girvin_exp}
\bibinfo{author}{\bibfnamefont{E.}~\bibnamefont{Chow}},
  \bibinfo{author}{\bibfnamefont{H.~P.} \bibnamefont{Wei}},
  \bibinfo{author}{\bibfnamefont{S.~M.} \bibnamefont{Girvin}},
  \bibinfo{author}{\bibfnamefont{W.}~\bibnamefont{Jan}}, \bibnamefont{and}
  \bibinfo{author}{\bibfnamefont{J.}~\bibnamefont{Cunningham}},
  \bibinfo{journal}{Phys. Rev. B} \textbf{\bibinfo{volume}{56}},
  \bibinfo{pages}{1676} (\bibinfo{year}{1997}).

\bibitem[{\citenamefont{Appleyard et~al.}(1998)\citenamefont{Appleyard,
  Nicholls, Simmons, Tribe, and Pepper}}]{appleyard_phonon}
\bibinfo{author}{\bibfnamefont{N.~J.} \bibnamefont{Appleyard}},
  \bibinfo{author}{\bibfnamefont{J.~T.} \bibnamefont{Nicholls}},
  \bibinfo{author}{\bibfnamefont{M.~Y.} \bibnamefont{Simmons}},
  \bibinfo{author}{\bibfnamefont{W.~R.} \bibnamefont{Tribe}}, \bibnamefont{and}
  \bibinfo{author}{\bibfnamefont{M.}~\bibnamefont{Pepper}},
  \bibinfo{journal}{Phys. Rev. Lett.} \textbf{\bibinfo{volume}{81}},
  \bibinfo{pages}{3491} (\bibinfo{year}{1998}).

\bibitem[{\citenamefont{Misguich et~al.}(1999)\citenamefont{Misguich,
  Lhuillier, Bernu, and Waldtmann}}]{misguich_ringexchange}
\bibinfo{author}{\bibfnamefont{G.}~\bibnamefont{Misguich}},
  \bibinfo{author}{\bibfnamefont{C.}~\bibnamefont{Lhuillier}},
  \bibinfo{author}{\bibfnamefont{B.}~\bibnamefont{Bernu}}, \bibnamefont{and}
  \bibinfo{author}{\bibfnamefont{C.}~\bibnamefont{Waldtmann}},
  \bibinfo{journal}{Phys. Rev. B} \textbf{\bibinfo{volume}{60}},
  \bibinfo{pages}{1064} (\bibinfo{year}{1999}).

\end{thebibliography}
\end{document}